\documentclass{svmult}

\usepackage{helvet}
\usepackage{courier}
\usepackage{type1cm}
\usepackage{amsmath}
\usepackage{amsfonts}
\usepackage{amssymb}
\usepackage{makeidx}         
\usepackage{graphicx}        
\usepackage{multicol}        
\usepackage[bottom]{footmisc}
\usepackage{chemarrow}
\usepackage{wasysym}

\makeindex             

\newcommand{\citep}[1]{\cite{#1}}

\begin{document}


\title*{Mathematical modeling of heterogeneous stem cell regeneration: from cell division to Waddington's epigenetic landscape}  

\titlerunning{Mathematical modeling of heterogeneous stem cell regeneration}

\author{Jinzhi Lei\orcidID{0000-0002-2670-6760}}

\vspace{0.5cm}

\institute{Jinzhi Lei \at School of Mathematical Sciences, Center for Applied Mathematics, Tiangong University, Tianjin 300387, China.
\email{jzlei@tiangong.edu.cn}}

\maketitle

\noindent\textbf{Abstract}\ \ 
Stem cell regeneration is a crucial biological process for most self-renewing tissues during the development and maintenance of tissue homeostasis.  In developing the mathematical models of stem cell regeneration and tissue development, cell division is the core process connecting different scale biological processes and leading to changes in cell population number and the epigenetic state of cells. This chapter focuses on the primary strategies for modeling cell division in biological systems. The Lagrange coordinate modeling approach considers gene network dynamics within each cell and random changes in cell states and model parameters during cell division. In contrast, the Euler coordinate modeling approach formulates the evolution of cell population numbers with the same epigenetic state via a differential-integral equation. These strategies focus on different scale dynamics, respectively, and result in two methods of modeling Waddington's epigenetic landscape: the Fokker-Planck equation and the differential-integral equation approaches. The differential-integral equation approach formulates the evolution of cell population density based on simple assumptions in cell proliferation, apoptosis, differentiation, and epigenetic state transitions during cell division. Moreover, machine learning methods can establish low-dimensional macroscopic measurements of a cell based on single-cell RNA sequencing data. The low dimensional measurements can quantify the epigenetic state of cells and become connections between static single-cell RNA sequencing data with dynamic equations for tissue development processes. The differential-integral equation presented in this chapter provides a reasonable approach to understanding the complex biological processes of tissue development and tumor progression.

\vspace{0.25cm}

\vspace{0.25cm}

\section{Introduction}

Biological processes are essentially multiscale dynamics at molecular, cellular, and tissue levels\cite{Eftimie:2016jq}. Individual cells usually show high heterogeneity at the molecular level with different expressions of genes and molecular interactions. At the cellular level, different types of cells undergo self-renew, differentiation, cell death, and migration; these behaviors are well-regulated to achieve a dynamic equilibrium in a tissue microenvironment. Biologically, the crosstalk between different scale dynamics is essential to maintain tissue growth. The multiscale biological processes occur at different spatial ranges and time scales. Mathematically, it is challenging to integrate the multiscale processes into a unified mathematical formulation to form a reasonable understanding of the biological process \cite{2007Natur.449..978W}.

Cell division is the core that connects the molecular and tissue-level development processes and maintains tissue homeostasis. In cellular regeneration, the genetic regulatory network underlying cell cycling regulates the decision between cell growth, differentiation, and cell division. The molecular level dynamics within a cell determine the signals that trigger the irreversible proliferation process. Moreover, cell divisions include molecular processes of DNA replication, epigenetic modifications reallocation/reconstruction, protein synthesis, and the partitioning of molecules at cell division. These processes are stochastic biochemical reactions that may result in variations in the newborn cells. At the macroscale level,  cell divisions result in changes in both the cell population and the density of different phenotypes of cells. Therefore, it is crucial to formulate cell division in mathematical models of biological systems in which microscale variations in each cell are considered.  

During the maintenance of tissue homeostasis, adult stem cells undergo cell divisions to replace dying cells and regenerate damaged tissues through controlled self-renewal and differentiation \cite{2015Sci...350.1319C}. Understanding the mechanisms that govern cell fate decisions and the regulation of self-renewal and differentiation in stem cells is of utmost significance. Waddington's epigenetic landscape has been fundamental in understanding cell fate decisions and differentiation \cite{Waddington:2012bu}. However, while the Waddington landscape provides an intuitive understanding of the biological process, the mechanisms driving cell fate decisions still need to be discovered \cite {Ferrell:2012fe, Hoppe2016, Rommelfanger2021}. Specifically, there are still debates on the mathematical formulations of Waddington landscapes associated with different types of biological processes \cite{Fard:2016aa, Li:2013aa, Feinberg:2023aa, Shi:2022gh}.    

This chapter reviews the mathematical models of cellular regeneration from micro to macro scales and introduces a general mathematical framework that integrates various biological processes involved at multiple scales. With the help of this framework, we propose a dynamic equation that describes the evolution of the Waddington landscape for the development of a multiple-cell system. 

Here, we briefly summarize the key formulations and ideas presented in this review.

At the microscale dynamics in a cell, the concentrations of functional molecules can be represented by a vector $\vec{x}$. The dynamics of the concentration of function molecules, $\vec{x}(t)$, is often described by a deterministic chemical rate equation of the form 
\begin{equation}
\label{eq:1.1}
\dfrac{\mathrm{d} \vec{x}}{\mathrm{d} t} = \vec{F}(\vec{x}).
\end{equation}
Alternatively, when noisy fluctuations are considered, we have a stochastic equation of the form
\begin{equation}
\label{eq:1.1-1}
\dfrac{\mathrm{d} \vec{x}}{\mathrm{d} t} = \vec{F}(\vec{x}) + \vec{\eta}(t),
\end{equation}
where $\vec{\eta}(t)$ represented the random fluctuation in the changes in molecule concentrations.

The function $\vec{F}$ describes the regulatory relationships among functional molecules, such as gene regulatory networks, protein-protein interactions, or modification of molecules. These regulations may affect the biochemical interactions involved in the molecules' production and degradation/dilution. The reaction rates are often represented by parameters $\vec{q}$ that are involved in the equation. Hence, the equation \eqref{eq:1.1} can be rewritten as (see section \ref{sec:2})
\begin{equation}
\label{eq:1.2}
\dfrac{\mathrm{d} \vec{x}}{\mathrm{d} t} = \vec{F}(\vec{x}; \vec{q})
\end{equation}
with parameter $\vec{q}$ explicitly included. Biologically, the parameter values $\vec{q}$ are usually non-constants and are dynamically changing over time.  Thus, the parameters $\vec{q}$ should be represented as $\vec{q}(t)$. 

During cell division, epigenetic modifications (histone modifications, DNA methylations, etc.) are redistributed and re-established at the daughter cells, followed by the reallocation of proteins and mRNA. During cell division, these processes result in discontinuous changes in variables $\vec{x}$ and parameters $\vec{q}$. Therefore, the above equation should be extended below over cell divisions
\begin{equation}
\label{eq:1.3}
\left\{
\begin{array}{ll}
\dfrac{\mathrm{d} \vec{x}}{\mathrm{d} t} = \vec{F}(\vec{x}; \vec{q}(t)), & \mbox{Between cell divisions}\\
(\vec{x}, \vec{q})\mapsto (\vec{x}', \vec{q}') \sim \mathcal{P}(\vec{x}', \vec{q}' | \vec{x}, \vec{q}), & \mbox{Cell division}
\end{array}
\right.
\end{equation}
Here, $\mathcal{P}(\cdot | \vec{x}, \vec{q})$ represents the random numbers with a conditional distribution that is dependent on $\vec{x}$ and $\vec{q}$. The equation of form \eqref{eq:1.3} provides a general framework for describing the dynamics of a single cell at a time scale across cell divisions. 

Cell behaviors such as cell division, apoptosis, and differentiation/aging must be considered to formulate the population dynamics of multiple cells across cell divisions. Moreover, we should also consider the heterogeneity of cells due to the variance in epigenetic states and the transition between epigenetic states during cell divisions. Let $Q(t, \vec{x})$ represent the number of cells with epigenetic state $\vec{x}$, and the evolution of $Q(t, \vec{x})$ can be modeled with the following differential-integral equation (see section \ref{sec:eul})
\begin{equation}
\label{eq:1.4}
\left\{
\begin{array}{rcl}
\dfrac{\partial Q(t, \vec{x})}{\partial t} &=& - Q(t, \vec{x}) (\beta(c(t), \vec{x})  + \kappa(\vec{x}))\\
&&{} \displaystyle+ 2 \int_\Omega \beta(c(t-\tau(\vec{y})), \vec{y}) Q(t - \tau(\vec{y}), \vec{y}) e^{-\mu(\vec{y})\tau(\vec{y}))} p(\vec{x}, \vec{y}) \mathrm{d} \vec{y},\\
c(t) &=&\displaystyle \int_\Omega Q(t, \vec{x}) \zeta(\vec{x}) \mathrm{d} \vec{x}.
\end{array}
\right.
\end{equation}
Here $\beta, \kappa, \mu$ represent the rates of proliferation, differentiation/senescence, and apoptosis, respectively; $\tau$ represents the duration of the proliferation phase; $\zeta$ represents the rate of cytokine secretion; $c$ stands for the concentration of growth factors secreted by all cells; $p(\vec{x}, \vec{y})$ quantifies the transition probability of epigenetic states during cell division (equation \eqref{eq:pxy} in section \ref{sec:eul}).

There are two methods to formulate the evolution of the Waddington landscape. Consider the microscale dynamics formulated by equation \eqref{eq:1.1-1}, where $\vec{\eta} = (\eta_1, \eta_2, \cdots, \eta_n)$ is a multi-dimensional Gaussian noise term, and the correlation satisfies $\langle \eta_i(t_1, \vec{x}) \eta_j(t_2, \vec{x}) \rangle = 2 D \delta_{i,j} \delta(t_1 - t_2)$, with $D$ as the diffusion coefficient. Let $P(t, \vec{x})$ represent the probability density of a cell in a state $\vec{x}$. Then, $P(t, \vec{x})$ satisfies the Fokker-Planck equation
\begin{equation}
\label{eq:fp1.0}
\dfrac{\partial P(t, \vec{x})}{\partial t}   = \nabla\cdot(D \nabla P - \vec{F} P).
\end{equation}
Moreover, if we introduce a birth-death rate $R(\vec{x})$ of a cell with state $\vec{x}$ and replace the probability density with the population density $f(t, \vec{x})$, the population density $f(t, \vec{x})$ satisfies the following population balance equation
\begin{equation}
\label{eq:bpe0}
\dfrac{\partial f}{\partial t} = \nabla \cdot (D \nabla f) - \nabla \cdot (f \vec{F}) + R f.
\end{equation}
The potential 
\begin{equation}
\label{eq:6}
U(t, \vec{x}) = - \log f(t, \vec{x})
\end{equation}
give a formulation of the Waddington landscape. Detailed discussions are given in section \ref{sec:wad}.

Alternatively, through equation \eqref{eq:1.4}, the evolution of the total cell number is given by
$$
Q(t) = \int_\Omega Q(t, \vec{x}) \mathrm{d} \vec{x},
$$
and the population density of cells with epigenetic state $\vec{x}$ is represented as
$$
f(t, \vec{x}) = \dfrac{Q(t, \vec{x})}{Q(t)}.
$$
From equation \eqref{eq:1.4}, the evolution equation of $f(t, \vec{x})$ is obtained as (equation \eqref{eq:fvh} in section \ref{sec:scra})
\begin{equation}
\label{eq:dvf}
\begin{aligned}
\dfrac{\partial f(t, \vec{x})}{\partial t} &= \dfrac{2}{Q(t)} \int_\Omega \beta(c_{\tau(\vec{y})}, \vec{y}) Q(t - \tau(\vec{y}), \vec{y}) e^{-\mu(\vec{y}) \tau(\vec{y})} (p(\vec{x}, \vec{y}) - f(t, \vec{x})) \mathrm{d} \vec{y} \\
&{}-f(t, \vec{x}) \int_\Omega f(t, \vec{y})\left((\beta(c, \vec{x}) + \kappa(\vec{x})) - (\beta(c, \vec{y}) + \kappa(\vec{y}))\right) \mathrm{d} \vec{y}.
\end{aligned}
\end{equation}
Equation \eqref{eq:dvf} describes the evolution dynamics of the population density, or the Waddington landscape according to equation \eqref{eq:6}, of a multicellular system when heterogeneity and plasticity are considered. 

The right-hand side of equation \eqref{eq:dvf} represents the growth of the population density $f(t, \vec{x})$. We denote it as the growth operator $\mathcal{R}[f]$. Referring to equation \eqref{eq:bpe0}, we can express this growth in the form of the population balance equation (see section \ref{sec:comb})
\begin{equation}
\label{eq:pbe1}
\dfrac{\partial f}{\partial t} = \nabla \cdot(D \nabla f) - \nabla \cdot (f \vec{F}) + \mathcal{R}[f].
\end{equation}  
Equation \eqref{eq:pbe1} provides a formulation for the evolution of the population density that integrates both molecular-level dynamics in a cell and cell population dynamics. The population dynamics $Q(t)$ and the population density $f(t, \vec{x})$ together provide macroscopic descriptions of the population dynamics of multiple cellular systems with cell heterogeneity and plasticity.  

\section{Modeling the cell cycle}
\label{sec:2}

The cell cycle is a highly regulated and orchestrated series of events that can be divided into four main phases: G1 (Gap 1), S (Synthesis), G2 (Gap 2), and M (Mitosis). During G1, the cell prepares for DNA synthesis; during the S phase, DNA replication occurs; and in G2, the cell prepares for mitosis. Finally, during the M phase, the cell undergoes mitosis, wherein the duplicated chromosomes are equally distributed between the two daughter cells. A complex network of regulatory proteins, checkpoints, and feedback loops tightly regulates the transition from one phase to another. 

Hundreds of mathematical models for the cell cycle have been published \cite{Ferrell:2011je}. These mathematical models describe the rate of change of different molecular species over time. Ordinary differential equations (ODEs) are the most common mathematical framework in cell cycling modeling. If $\vec{x}$ represents the concentrations of particular molecules involved in the cell-cycling regulation circuit, the ODE would be expressed as 
\begin{equation}
\label{eq:1.8}
\dfrac{\mathrm{d} \vec{x}}{\mathrm{d} t} = \vec{F}(\vec{x}),
\end{equation}
the function $\vec{F}$ is determined by how the molecules interact with each other to form autonomous oscillators in cell cycling.  

For instance, a simple protein circuit in the in \textit{Xenopus} embryos' cell cycle centers on cyclin-dependent protein kinase (CDK1), the anaphase-promoting complex (APC), and a protein like Polo-like kinase 1 (Plk1). In this circuit, CDK1 activates APC through Plk1, and APC inactivates CDK1 to form negative feedback. Let $[\mathrm{CDK1}^*]$, $[\mathrm{APC}^*]$, and $[\mathrm{Plk1}^*]$ represent the concentrations of active CDK1, APC, and Plk1, respectively.  The interactions described result in the following two ODEs model \cite{Ferrell:2011je}:
\begin{equation}
\label{eq:1.9}
\left\{
\begin{array}{rcl}
\displaystyle \dfrac{\mathrm{d} [\mathrm{CDK1}^*]}{\mathrm{d} t} &=& \displaystyle a_1  - b_1 [\mathrm{CDK1}^*] \dfrac{[\mathrm{APC}^*]^{n_1}}{K_1^{n_1} +  [\mathrm{APC1}^*]^{n_1}},\\
\displaystyle \dfrac{\mathrm{d} [\mathrm{Plk1}^*]}{\mathrm{d} t} &=& \displaystyle a_2 (1 - [\mathrm{Plk1}^*]) \dfrac{[\mathrm{CDK1}^*]^{n_2}}{K_2^{n_2} + [\mathrm{CDK1}^*]^{n_2}} - b_2 [\mathrm{Plk1}^*],\\
\displaystyle \dfrac{\mathrm{d} [\mathrm{APC}^*]}{\mathrm{d} t}&=& \displaystyle a_3 (1 - [\mathrm{APC}^*]) \dfrac{[\mathrm{Plk1}^*]^{n_3}}{K_3^{n_3} + [\mathrm{Plk1}^*]^{n_3}}  - b_3 [\mathrm{APC}^*]. 
\end{array}
\right.
\end{equation}
Here, the parameters $a_i$ represent the maximum protein production/activation rates, while $b_i$ represent the degradation/inactivation rates. The total concentrations of active and inactive APC and Plk1 are assumed to be constant and normalized to 1. Equation \eqref{eq:1.9} provides an example of equation \eqref{eq:1.8} with $\vec{x} = ([\mathrm{CDK1}^*], [\mathrm{APC}^*], [\mathrm{Plk1}^*])$, and $\vec{F}$ is defined by the right-hand side of \eqref{eq:1.9}. Sustained oscillatory dynamics can form with a proper selection of model parameters. 

In cell-cycle modeling, we often consider constant model parameters for simplicity. However, the model parameters change with time for different reasons, such as cell growth and extracellular perturbations. Thus, the regulation function $\vec{F}$ often depends on time-dependent parameters $\vec{q}(t)$, and the equation of form \eqref{eq:1.8} can be expressed as
\begin{equation}
\label{eq:1.10}
\dfrac{\mathrm{d} \vec{x}}{\mathrm{d} t} = \vec{F}(\vec{x}; \vec{q}).
\end{equation}
For instance, the gene expression rates $a_i, (i=1,2,3)$ may be subjected to fluctuations in the intracellular microenvironment and epigenetic modification, and hence $\vec{q} = (a_1, a_2, a_3)$. Given the initial condition $\vec{x}(t_0) = \vec{x}_0$ and the temporal dynamics $\vec{q}(t)$, we can solve the equation \eqref{eq:1.10} to obtain the time course of the system state $\vec{x}(t)$ for $t > t_0$. 

The above time course is valid only within one cell cycle, and the dynamics across cell division must be considered to describe the long-term dynamics.   Discontinuous changes in molecule concentration may happen following mitosis, by which molecules in a cell are redistributed to the two daughter cells. Moreover, epigenetic markers, including histone modifications and DNA methylations, are re-established for the newborn cells following DNA replication during the S phase of cell cycling. Thus, following cell divisions, a mother cell divides into two daughter cells, both system state $\vec{x}$ and parameter $\vec{q}$ may undergo discontinuous changes at the end of mitosis. We assume that the state $\vec{x}$ and the parameter $\vec{q}$ change following random processes. Hence, their values after mitosis should be taken as random numbers following a conditional distribution that depends on their mother cell values. The conditional distribution is denoted as $\mathcal{P}(\cdot | \vec{x}, \vec{q})$. Thus, we can extend equation \eqref{eq:1.10} to a discontinuous dynamical equation across cell division as
\begin{equation}
\label{eq:1.11}
\left\{
\begin{array}{ll}
\dfrac{\mathrm{d} \vec{x}}{\mathrm{d} t} = \vec{F}(\vec{x}; \vec{q}), & \mbox{Between cell divisions}\\
(\vec{x}, \vec{q})\mapsto (\vec{x}', \vec{q}') \sim \mathcal{P}(\vec{x}', \vec{q}' | \vec{x}, \vec{q}), & \mbox{Cell division}
\end{array}
\right.
\end{equation}

Biologically, random changes in the state $\vec{x}$ are often associated with the random partition of molecules at cell division \cite{2011PNAS..10815004H}. Let $V_d$ and $V_b$ be the cell volume at division and newborn, respectively. For the component $x_i$, the total number of $N_i = V_d x_i$ molecules are randomly partitioned to two daughter cells, and the molecule number at one newborn cell is $V_b x_i'$.  Let $r_i = V_b x_i'/V_d x_i$ represent the fraction of molecules allocated to one daughter cell, then $0< r_i \leq  1$. Mathematically, the distribution of molecules partitioned into two daughter cells can be characterized by a binomial distribution random number, with the partition rate following a conjugate prior distribution--the beta distribution. Consequently, we can assume that $r_i$ follows a beta distribution, providing a rule for how $x_i$ changes during cell division.

Epigenetic regulations, such as histone modifications or DNA methylations, can interfere with chromatin structure and alter the gene expression rates. Hence, the expression rates $a_i$ depend on each gene's epigenetic modification state $u_i$. The epigenetic state of each gene can refer to the fractions of marked nucleosomes or methylated CpG sites in the DNA segment of interest. Since the epigenetic states primarily affect the chromatin structure, they might influence the chemical potential required to initiate the transcription process. Thus, the expression rates $a_i$ in \eqref{eq:1.9} can be expressed as
$$
a_i = \alpha_i e^{\lambda_i u_i},\quad i=1, 2, 3.
$$ 
Here, $\lambda_i$ represents the impact of the epigenetic modifications on the expression rates. Specifically, $\lambda_i > 0$ indicates an epigenetic modification that enhances the strength of cell activation, while $\lambda_i <0$ indicates a modification that reduces this strength. The epigenetic state $u_i$ may experience random changes during cell divisions, leading to corresponding random fluctuations in the expression rates $a_i$. For example, when $u_i$ denotes the fraction of marked nucleosomes or methylated CpG sites, it lies within the $0<u_i<1$ range. Following a similar rationale as discussed for the distribution of $r_i$, we can posit that the value of $u_i$ follows a beta distribution, with shape parameters depending on the cell state before cell division.

Given the rule of how $\vec{x}$ and $\vec{q}$ may change at mitosis, solving equation \eqref{eq:1.11} gives the time course $(\vec{x}(t), \vec{q}(t))$ across multiple cell cycles, which corresponds to long-term tracking of a cell over numerous cell cycles simulation. Nevertheless, \eqref{eq:1.11} only describes the microscale dynamics inside a cell and cannot model the population dynamics of multiple cells.

\section{Modeling homogeneous stem cell regeneration}

\begin{figure}[htbp]
\centering
\includegraphics[width=8cm]{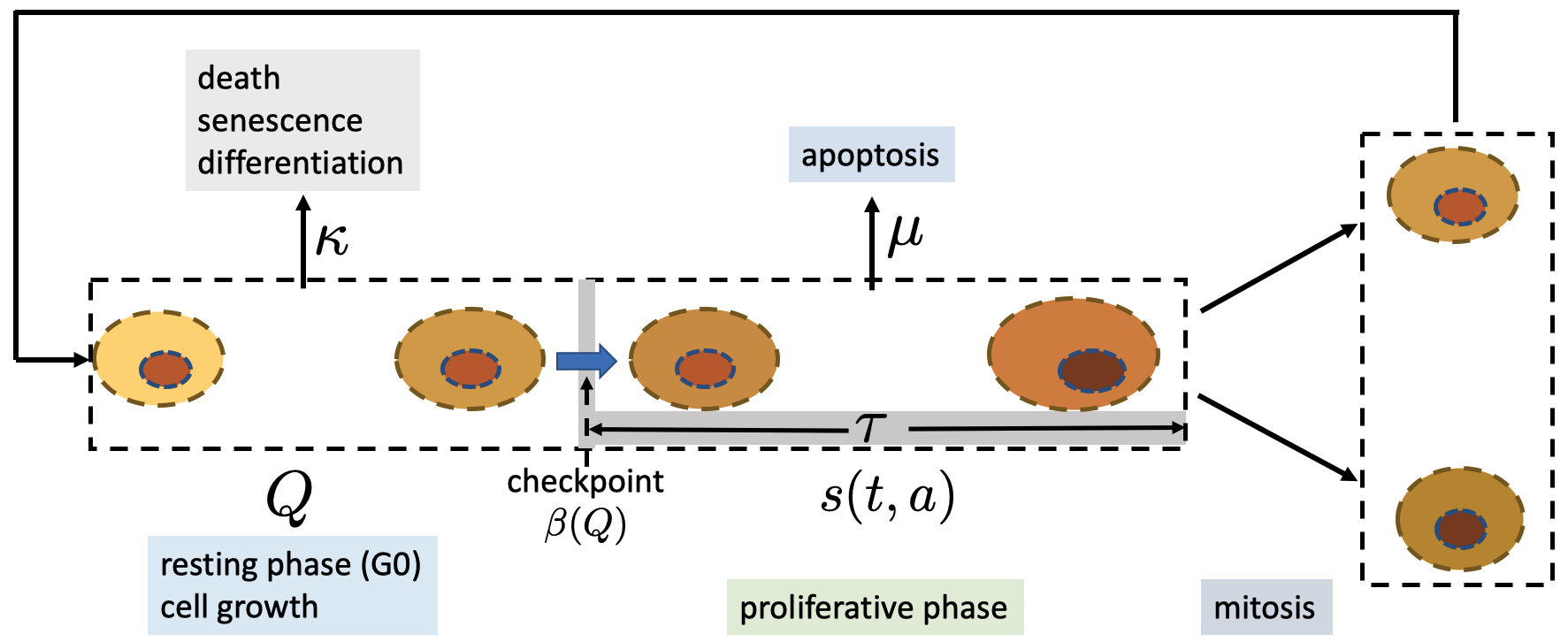}
\caption{The G0 model of stem cell regeneration. Here, $Q$ represents the number of cells in the resting phase, and $s(t, a)$ represents the number of cells at time $t$ with age $a$ in the proliferating phase. Stem cells in the resting phase may enter the proliferative with a rate $\beta$ or can be removed from the resting pool with a rate $\kappa$. The cells in the proliferating phase are randomly lost at a rate $\mu$ or undergo mitosis at a time $\tau$ after entering the proliferative compartment.}
\label{fig:3.1}
\end{figure}

It is crucial to understand how a system consisting of multiple cells evolves. To study the dynamics of stem cell regeneration, we employ the G0 cell cycle model. According to this model, cells go through a resting phase called G0, where they grow and prepare to enter the proliferative phase when they receive cell cycling checkpoint signals. Stem cells in the resting phase may enter the proliferative phase at a rate $\beta$, which incorporates negative feedback to signaling response pathways, or they can be removed from the resting pool at a rate $\kappa$ through the biological processes such as differentiation, senescence, or death. The cells in the proliferating phase are randomly lost at a rate $\mu$ or undergo mitosis at a fixed time $\tau$ after entering the proliferative compartment. Each mother cell generated two daughter cells at mitosis. The newborn cells enter the resting phase to start the next cycle. This process is represented in Figure \ref{fig:3.1}.

During the resting phase, changes in cell numbers can be described using ordinary differential equations. However, during the proliferating phase, the biological processes can be modeled using the age-structured model, also known as the transport equation introduced by M'kendrick in 1925 for medical problems \cite{MKendrick:1925}. The age-structured stem cell dynamics model given below was first proposed by Burns and Tannock in 1970 \cite{Burns:1970tm}. Later, in 1978, Michael Mackey used this model to study periodic hematopoiesis \cite{Mackey:1978vy}.

We assume that all cells have identical micro-scale properties and are only interested in the time course of cell numbers. Let $Q(t)$ be the number of cells in the resting phase, and $s(t, a)$ be the number of stem cells at time $t$ with age $a$ in the proliferating phase. The above biological process can be described by the following age-structured equation \citep{Burns:1970tm,Mackey:1978vy,Jinzhi:2011bl}
\begin{equation}
\label{eq:9}
\begin{aligned}
&\dfrac{\partial s(t, a)}{\partial t} + \dfrac{\partial s(t, a)}{\partial a} = - \mu s(t, a), \quad t > 0,\, 0 < a <\tau\\
&\dfrac{\mathrm{d} Q}{\mathrm{d} t} = 2 s(t, \tau)  - (\beta(Q) +\kappa) Q, \quad t > 0.
\end{aligned}
\end{equation}
The boundary condition at age $a = 0$ is given by
\begin{equation}
s(t, 0) = \beta(Q(t)) Q(t).
\end{equation}
Here, the proliferation rate of resting phase cells is represented by the function $\beta(Q)$, which depends on the number of cells in the resting phase.

In \eqref{eq:9}, $\kappa$ signifies the rate at which cells irreversibly transition out of the resting phase. This transition may result from various biological processes, such as differentiation, senescence, or cell death. For conciseness, we will henceforth denote $\kappa$ as the ``differentiation rate.''

The first equation in \eqref{eq:9} can be integrated through the method of characteristic line, which results in a delay differential equation
\begin{equation}
\label{eq:5.15}
\dfrac{\mathrm{d} Q}{\mathrm{d} t} =  - (\beta(Q) + \kappa) Q + 2 e^{-\mu \tau}\beta(Q_\tau) Q_\tau,
\end{equation}
where $Q_\tau(t) = Q(t-\tau)$. This equation describes the general population dynamics of stem cell regeneration.

Here, the lag time $\tau$ is automatically introduced following the age-structured equation in \eqref{eq:9}, and $\tau$ represents the duration of the proliferation phase. If the lag time $\tau$ is omitted,  \eqref{eq:5.15} becomes an ODE model
\begin{equation}
\dfrac{\mathrm{d} Q}{\mathrm{d} t} = (2 e^{-\mu}-1) \beta(Q) Q - \kappa Q.
\end{equation}
Here, we note that by omitting the delay, we cannot simply set $\tau = 0$ in \eqref{eq:5.15}, but replace the term $e^{-\mu \tau}$ with $e^{-\mu}$ that represents the survival probability of a cell after the proliferation phase.

Biologically, the self-renewal ability of a cell is intricately linked to microenvironmental conditions, such as \textit{e.g.} growth factors and various types of cytokines, as well as intracellular signaling pathways \citep{Massague:2012bd,Moustakasetal.2002,Yangetal.2010,Ornitz:2001vh}. Despite the complexity of signaling pathways, the phenomenological formation of Hill function dependence can be derived from simple assumptions regarding the interactions between signaling molecules and receptors \citep{Bernard:2003ct,Lei:2020fw}. 

For instance, we assume that the niche secretes the positive growth factors and that the cells release growth factor inhibitors.  Different types of cytokines bind to cell surface receptors to regulate cell behavior. Let $[\mathrm{L}]$ denote the concentration of ligands for growth factor inhibitor; $[\mathrm{R}]$ denote the density of free receptor; $[\mathrm{R^*}]$ denotes the density of activated receptors, and $Q$ denotes the stem cell number. The total number of receptors is 
\begin{equation}
\label{eq:bet1}
[\mathrm{R}] + [\mathrm{R}^*] = m Q. 
\end{equation}
At the equilibrium, we have the following equation
\begin{equation}
\label{eq:bet2}
[\mathrm{R}] [\mathrm{L}]^n = K [\mathrm{R}^*],
\end{equation}
where $K$ is the equilibrium constant. We assume that the active receptors inhibit cell proliferation, and hence the proliferation rate $\beta$ is proportional to the fraction of free receptors on a cell, \textit{i.e.},
$$
\beta = \beta_0 \dfrac{[\mathrm{R}]}{m Q}.
$$ 
From equations \eqref{eq:bet1} and \eqref{eq:bet2}, we obtain the fraction of free receptors:
$$
\dfrac{[\mathrm{R}]}{m Q}  = \dfrac{K}{K + [\mathrm{L}]^n}.
$$
When the ligands are secreted from stem cells and cleaned constantly, the ligand concentration is proportional to the cell number, resulting in $[\mathrm{L}] = \sigma Q$. These calculations lead to the following Hill-type function of the proliferation rate:  
\begin{equation}
\label{eq:6.23-1}
\beta(Q) = \beta_0 \dfrac{\theta^{m_0}}{\theta^{m_0} + Q^{m_0}},
\end{equation}
where $\beta_0$ represents the maximum proliferation rate of normal cells, and $\theta=\sqrt{n}{K}/\sigma$ is a constant for the half-effective cell number.

Moreover, considering cancer cells that may exhibit uncontrolled cell growth, an extra factor $\beta_1$ can be introduced, leading to 
\begin{equation}
\label{eq:6.23}
\beta(Q) = \beta_0 \dfrac{\theta^{m_0}}{\theta^{m_0} + Q^{m_0}} + \beta_1,
\end{equation} 
where the positive parameter $\beta_1$ accounts for potential mutations in cancer cells that enable sustained proliferative signaling or evasion of growth suppressors, representing a hallmark of cancer \citep{Hanahan:2000hx}.  Physical interactions can restrict cell growth within a more realistic model, causing $\beta_1$ to decrease to zero as the cell number $Q$ becomes sufficiently large.

From \eqref{eq:5.15}, the steady state $Q(t) \equiv Q^*$ is given by the equation
$$
- (\beta(Q^*)  + \kappa) Q^* +2 e^{-\mu \tau} \beta(Q^*) Q^* = 0,
$$
which yields either $Q^* = 0$, or
\begin{equation}
\label{eq:bs}
\beta(Q^*) = \dfrac{\kappa}{2 e^{-\mu \tau} - 1}.
\end{equation}
When $\beta(Q)$ is given by \eqref{eq:6.23}, the equation \eqref{eq:5.15} has a unique positive steady state if and only if 
$$
\beta_0 > \dfrac{\kappa}{2 e^{-\mu \tau} - 1} - \beta_1 > 0.
$$
In particular, when
\begin{equation}
\label{eq:6.22}
\beta_1 \geq \dfrac{\kappa}{2 e^{-\mu \tau} - 1},
\end{equation}
zero solution $Q \equiv 0$ is the only steady state and is unstable. 

Let $\bar{Q} = 1/Q$, and we have the equation for $\bar{Q}(t)$:
$$
\dfrac{\mathrm{d} \bar{Q}}{\mathrm{d} t} = (\beta(\bar{Q}^{-1})+\kappa)\bar{Q} + 2 e^{-\mu \tau} \beta(\bar{Q}_\tau^{-1}) (\bar{Q}/\bar{Q}_\tau) \bar{Q}.
$$
When \eqref{eq:6.22} is satisfied, the zero solution $\bar{Q}\equiv 0$ is stable. Consequently, all positive solutions of the original equation \eqref{eq:5.15} approach infinity when $t\to\infty$\footnote{A formal mathematical proof of this statement remains open.}, indicating uncontrolled growth. Therefore, the inequality \eqref{eq:6.22} summarizes a  general condition for uncontrolled growth, \textit{i.e.}, malignant tumors. The inequality is satisfied when $\beta_1$ is increased, along with the decrease of $\mu$ and $\kappa$.  Biologically, these conditions correspond to self-sufficiency in growth, insensitivity to antigrowth signals, evasion of apoptosis, and dysregulation in the differentiation (or senescence) pathways. These are well-known hallmarks of cancer \citep{Hanahan:2000hx}. Hence, a simple analysis for the homogeneous stem cell regeneration model \eqref{eq:5.15} can reveal key hallmarks of cancer.

\section{Modeling cellular heterogeneity}

The equation \eqref{eq:1.11} explains the time course of a single cell without considering population dynamics. On the other hand, \eqref{eq:5.15} describes population dynamics but doesn't provide microscale information of individual cells. To understand stem cell regeneration with cellular heterogeneity, we need to combine microscale fluctuation with macroscale population dynamics. This can be achieved by applying the mathematical modeling framework of either Lagrange or Euler coordinates similar to fluid dynamics.       

\subsection{Lagrange coordinate modeling}

Based on the Lagrange coordinate modeling, we consider each cell individually. Thus, let $\Sigma_t = \{[C_i]_{i=1}^{Q(t)}\}$ be the collection of all cells at time $t$, where $Q(t)$ represents the number of cells and $C_i = (\vec{x}_i(t), \vec{q}_i(t))$. We extend the equation \eqref{eq:1.11} to include all cells
\begin{equation}
\label{eq:1.17}
\left\{
\begin{array}{ll}
\dfrac{\mathrm{d} \vec{x}_i}{\mathrm{d} t} = \vec{F}(\vec{x}_i; \vec{q}_i), & \mbox{Between cell divisions}\\
(\vec{x}_i, \vec{q}_i)\mapsto (\vec{x}'_i, \vec{q}'_i) \sim \mathcal{P}(\vec{x}'_i, \vec{q}'_i | \vec{x}_i, \vec{q}_i), & \mbox{Cell division}
\end{array}
\right.
\end{equation}
Here, each cell corresponds to a set of differential equations. The cell number $Q(t)$ changes with time due to cell death and cell division, so the total number of equations in \eqref{eq:1.17} varies over time. Moreover, the cells do not divide synchronously in the equation \eqref{eq:1.17}; hence, it is challenging to write down a unified equation for all cells. 

The Lagrange coordinate model \eqref{eq:1.17} describes the microscale dynamics inside each cell; however, it is mathematically difficult to formulate and study. In numerical studies, we can apply the agent-based modeling technique to simulate the dynamics of multiple cells.   
  
\subsection{Euler coordinate modeling}
\label{sec:eul}

To model cellular heterogeneity within the framework of Euler coordinate modeling, we introduce a variable $\vec{x}$ (often a high-dimensional vector) for the epigenetic state of a cell and $\Omega$ for the space of all possible epigenetic states in resting phase stem cells \cite{Lei:2020fw,2014PNAS..111E.880L,Lei:2020ei}. The epigenetic state $\vec{x}$ represents intrinsic cellular states that may dynamically change over time, whether in a cell cycle or during cell division. Biologically, the epigenetic state of a cell can be any molecular level changes that are independent of the DNA sequences, including the patterns of DNA methylation, nucleosome histone modifications, and transcriptomics   \cite{Probst:2009iq,Schepeler:2014ir,SerraCardona:2018bz,Singer:2014eua,Takaoka:2014gg,Wu:2014gw}.

Through the epigenetic state $\vec{x}\in \Omega$, let $Q(t, \vec{x})$ represent the number of cells at time $t$ in the resting phase and with epigenetic state $\vec{x}$.  Now, the total cell number is given by
\begin{equation}
\label{eq:Nt}
Q(t) = \int_{\Omega} Q(t, \vec{x}) \mathrm{d} \vec{x}.
\end{equation}

The proliferation of each cell is regulated by the signaling pathways that are dependent on extracellular cytokines released by all cells in the niche and the epigenetic state $\vec{x}$ of the cell \cite{Bernard:2003ct,Lander:2009fr,Mangel:2012ct}. Let $\zeta(\vec{x})$ be the rate of cytokine secretion by a cell with state $\vec{x}$, and
\begin{equation}
c(t) = \int_{\Omega} Q(t, \vec{x}) \zeta(\vec{x}) \mathrm{d} \vec{x}
\end{equation}
represents the effective concentration of cytokines to regulate cell proliferation. Similar to \eqref{eq:6.23}, the proliferation rate $\beta$ can be written as a function of cytokine concentration $c$ and the epigenetic state $\vec{x}$, \textit{i.e.},
\begin{equation}
\label{eq:beta}
\beta(c, \vec{x}) = \beta_0(\vec{x}) \dfrac{\theta(\vec{x})^{m_0}}{\theta(\vec{x})^{m_0} + c^{m_0}} + \beta_1(\vec{x}).
\end{equation}
Moreover, the apoptosis rate $\mu$, the cell cycle duration $\tau$, and the differentiation rate $\kappa$ are dependent on the epigenetic state $\vec{x}$ and are denoted by $\mu(\vec{x})$, $\tau(\vec{x})$, and $\kappa(\vec{x})$, respectively.  We assume that these rates depend solely on the state of each cell without considering the cell-to-cell interactions. 

During cell division, a single mother cell splits into two daughter cells. However, the daughter cells may not share the same epigenetic state as the mother cell, resulting in cell plasticity during cellular regeneration. To account for this plasticity, we introduce an inheritance function (\textit{aka} transition function), denoted by $p(\vec{x}, \vec{y})$, which represents the probability that a daughter cell with state $\vec{x}$ has originated from a mother cell with state $\vec{y}$ after cell division, \textit{i.e.}, the conditional probability density
\begin{equation}
\label{eq:pxy}
p(\vec{x}, \vec{y}) = P(\mbox{state of daughter cell }= \vec{x}\ \big|\ \mbox{state of mother cell } = \vec{y}).
\end{equation}
The inheritance function is used to consider cell plasticity during each cell cycle. It is obvious that
$$\int_\Omega p(\vec{x}, \vec{y}) \mathrm{d} \vec{x} = 1$$
for any $\vec{y}\in \Omega$.

Now, similar to \eqref{eq:9}, when stem cell heterogeneity is included, we obtain the corresponding age-structured model equation
\begin{equation}
\label{eq:m14}
\begin{array}{rcl}
\displaystyle \nabla' s(t, a, \vec{x}) &=& \displaystyle - \mu(\vec{x}) s(t, a,\vec{x}), \quad (0 < a < \tau(\vec{x}))\\
\displaystyle \dfrac{\partial Q(t, \vec{x})}{\partial t} &=& \displaystyle 2 \int_\Omega s(t, \tau(\vec{y}), \vec{y})p(\vec{x}, \vec{y}) \mathrm{d} \vec{y} - (\beta(c(t),\vec{x}) + \kappa(\vec{x})) Q(t, \vec{x}),
\end{array}
\end{equation}
and
$$s(t,0,\vec{x}) = \beta(c(t), \vec{x})Q(\vec{x}, t),\quad c(t) = \int_\Omega Q(t, \vec{x}) \zeta(\vec{x}) \mathrm{d} \vec{x}.$$
Here, $\nabla' = \partial/\partial t + \partial/\partial a$ represents the age-structured operator, and the epigenetic state $\vec{x}$ can be considered a parameter for the first equation. We can apply the characteristic line method to solve the first equation of \eqref{eq:m14} and obtain
$$s(t,\tau(\vec{x}), \vec{x}) = \beta(c(t-\tau(\vec{x})), \vec{x}) Q(t-\tau(\vec{x}), \vec{x})e^{-\mu(\vec{x}) \tau(\vec{x})}.$$
Thus, substituting $s(t, \tau(\vec{x}), \vec{x})$ into the second equation in \eqref{eq:m14}, we obtain the following delay differential-integral equation (here, we only show the equation for $t\geq \tau$ that is important for the long-term behavior)
\begin{equation}
\label{eq:6.26}
\left\{
\begin{array}{rcl}
\displaystyle\dfrac{\partial Q(t,\vec{x})}{\partial t} &=& \displaystyle-Q(t, \vec{x}) (\beta(c,\vec{x})  + \kappa(\vec{x}))\\
&&\displaystyle{} + 2 \int_{\Omega} \beta(c(t-\tau(\vec{y})),\vec{y}) Q(t-\tau(\vec{y}),\vec{y}) e^{-\mu(\vec{y})\tau(\vec{y})} p(\vec{x},\vec{y}) \mathrm{d}\vec{y},\\
c(t)&=&\displaystyle\int_{\Omega} Q(t,\vec{x}) \zeta(\vec{x}) \mathrm{d}\vec{x}.
\end{array}
\right.
\end{equation}

Equation \eqref{eq:6.26} provides a general mathematical framework for modeling the dynamics of heterogeneous stem cell regeneration with the epigenetic transition. 

From \eqref{eq:6.26}, and integrating both sides of the equation, we obtain
\begin{equation}
\label{eq:sfg}
\begin{aligned}
\dfrac{\mathrm{d} Q}{\mathrm{d} t} = &-\int_\Omega Q(t, \vec{x}) (\beta(c, \vec{x}) + \kappa(\vec{x})) \mathrm{d} \vec{x}\\
&{} + 2 \int_\Omega \beta(c_{\tau(\vec{x})},\vec{x}) Q(t - \tau(\vec{x}), \vec{x}) e^{-\mu(\vec{x})\tau(\vec{x})} \mathrm{d} \vec{x}.
\end{aligned}
\end{equation}
If we omit the heterogeneity, all rate functions are independent of $\vec{x}$ and $c(t) = Q(t)$, we re-obtain the delay differential equation \eqref{eq:5.15} for homogeneous stem cell regeneration.

Biologically, equation \eqref{eq:6.26} connects different scale components: the gene expression values at the single-cell level ($\vec{x}$), the population dynamic properties ($\beta(c,\vec{x}), \kappa(\vec{x})$, and $\mu(\vec{x})$), cell cycle ($\tau(\vec{x})$), cytokine secretion ($\zeta(\vec{x})$), and the transition of epigenetic states ($p(\vec{x}, \vec{y})$). In this equation, the functions $\beta(c, \vec{x})$, $\kappa(\vec{x})$, $\mu(\vec{x})$, $\tau(\vec{x})$ describe the kinetic properties of cell cycling and are termed as the \textit{kinetotype} of a cell \citep{Lei:2020fw}. This framework can be applied to different problems related to cell regeneration, such as development, aging, and tumor evolution \cite{Lei:2020fw,Lei:2020ei,Zhang:2021fs,Liang:2023aa}.

For mathematical simplicity, we can omit the delay $\tau$ in equation \eqref{eq:6.26}. It is important to note that the delay originates from the duration of the proliferative phase. Therefore, when the delay is omitted, we set $\tau(\vec{y}) = 0$ in $c(t-\tau(\vec{y}))$ and $Q(t-\tau(\vec{y}), \vec{y})$. Additionally, we replace $e^{-\mu(\vec{y}) \tau(\vec{y})}$ with $e^{-\mu(\vec{y})}$ to represent the survival rate of cells during the proliferative phase. Consequently, when the delay is omitted, equation \eqref{eq:6.26} becomes
\begin{equation}
\label{eq:6.26-1}
\left\{
\begin{array}{rcl}
\displaystyle\dfrac{\partial Q(t,\vec{x})}{\partial t} &=& \displaystyle-Q(t, \vec{x}) (\beta(c,\vec{x})  + \kappa(\vec{x}))\\
&&\displaystyle{} + 2 \int_{\Omega} \beta(c(t),\vec{y}) Q(t,\vec{y}) e^{-\mu(\vec{y})} p(\vec{x},\vec{y}) \mathrm{d}\vec{y},\\
c(t)&=&\displaystyle\int_{\Omega} Q(t,\vec{x}) \zeta(\vec{x}) \mathrm{d}\vec{x}.
\end{array}
\right.
\end{equation}
The discussions hereafter also apply to this equation when the delay is omitted. 

The mathematical framework presented in equation \eqref{eq:6.26} offers a basic model that incorporates the fundamental components of stem cell regeneration, such as cell cycling, cellular heterogeneity, and plasticity. However, stem cell systems in biological processes can be much more complex, and additional biological processes should be incorporated into the framework. For instance, the gene networks that underlie the dynamics of epigenetic states within a cell cycle, cell-to-cell interactions within a niche, and the interaction between all cells and the microenvironmental factors.

In the model given by \eqref{eq:6.26}, the variable $\vec{x}$ signifies the epigenetic state of a cell during the resting phase. For simplicity, we disregard changes in the epigenetic state within the resting phase. The model exclusively considers variations in the epigenetic state during cell division, encapsulated by the inheritance function $p(\vec{x}, \vec{y})$. However, this simplicity excludes the differentiation of cells during the resting phase, a departure from some biological observations indicating that differentiation can occur independently of cell division \cite{ONeill:1972kg,Li:2014jm,Reilein:2018gd,Muhr:2021bd,Burda:2022fd,Kukreja:2023gd}.

Several mathematical models have been proposed to describe division-independent differentiation, often formulated as a series of ordinary differential equations for discrete cell lineages or maturation stages \cite{Adimy:2006,Glauche:2007gu,Lander:2009fr,Adimy:2014gs}. In our model, division-independent differentiation is represented by the differentiation rate $\kappa(\vec{x})$. It is possible to extend the model equation \eqref{eq:6.26} to include discrete cell lineages, resulting in the formulation
\begin{equation}
\label{eq:25}
\begin{array}{rcl}
\displaystyle\dfrac{\partial Q_i(t,\vec{x})}{\partial t} &=& \displaystyle\kappa_{i-1}(\vec{x}) Q_{i-1}(t, \vec{x}) -Q_i(t, \vec{x}) (\beta_i(c_i,\vec{x})  + \kappa_i(\vec{x}))\\
&&\displaystyle{} + 2 \int_{\Omega} \beta_i(c_i(t-\tau_i(\vec{y})),\vec{y}) \\
&&\displaystyle{} \qquad\qquad\qquad \times Q_i(t-\tau_i(\vec{y}),\vec{y}) e^{-\mu_i(\vec{y})\tau_i(\vec{y})} p_i(\vec{x},\vec{y}) \mathrm{d}\vec{y},\\
c_i(t) &=&\displaystyle \int_\Omega Q_i(t, \vec{x})\zeta_i(\vec{x}) \mathrm{d} \vec{x}.
\end{array}
\end{equation}
Here, the subscript $i$ denotes the $i$'th subtype of cells, along with their cell lineage or maturation stage. Mathematically, the differentiation represented by $\kappa$ is one-directional and irreversible, while cell plasticity, described by the inheritance function $p(\vec{x}, \vec{y})$, is multi-directional and reversible.

In a manner akin to the extension presented in \eqref{eq:25}, we can further extend the equation \eqref{eq:6.26} to encompass gene mutations. Gene mutations typically occur in one direction following DNA replication during cell divisions. Consequently, we can introduce discrete mutant types and a mutation rate matrix $(p_{i,j}(\vec{x}))$, where $p_{i,j}$ denotes the mutation rate that changes the mutant type $j$ to type $i$. The mutation rates may depend on the epigenetic state $\vec{x}$. The formulation involving gene mutation can be expressed as
\begin{equation}
\label{eq:27}
\begin{array}{rcl}
\displaystyle\dfrac{\partial Q_i(t,\vec{x})}{\partial t} &=& \displaystyle  -Q_i(t, \vec{x}) (\beta_i(c,\vec{x})  + \kappa_i(\vec{x}))\\
&&\displaystyle{} + 2 \int_{\Omega} (1 - \sum_{j\not= i}p_{j,i}(\vec{y}))\beta_i(c(t-\tau_i(\vec{y})),\vec{y})\\
&&\displaystyle{}\qquad\qquad\qquad\times Q_i(t-\tau_i(\vec{y}),\vec{y}) e^{-\mu_i(\vec{y})\tau_i(\vec{y})} p_i(\vec{x},\vec{y}) \mathrm{d}\vec{y},\\
&&\displaystyle{} + 2 \sum_{j\not= i} \int_{\Omega} p_{i,j}(\vec{y})\beta_j(c(t-\tau_j(\vec{y})),\vec{y})\\
&&\displaystyle{}\qquad\qquad\qquad\times  Q_j(t-\tau_j(\vec{y}),\vec{y}) e^{-\mu_j(\vec{y})\tau_j(\vec{y})} p_j(\vec{x},\vec{y}) \mathrm{d}\vec{y}\\
c(t) &=&\displaystyle \sum_{i} \int_\Omega Q_i(t, \vec{x})\zeta_i(\vec{x}) \mathrm{d} \vec{x}.
\end{array}
\end{equation}
Please refer to \cite{Lei:2020fw} for examples and further discussions.

Please note that the term ``stem cell'' used in this context has a different meaning than its biological use. In organisms, stem cells are either undifferentiated or partially differentiated cells that can differentiate into various cell types and replicate indefinitely. Their self-renewal ability through cell division and their capacity to differentiate into specialized cell types distinguish stem cells from progenitor cells that cannot divide indefinitely and precursor or blast cells that are typically committed to differentiating into a specific cell type.

However, the mathematical models introduced here mainly focus on cell population dynamics without considering cell types. Therefore, the models do not explicitly include stem cell differentiation and cell type transitions. The model considers cells that can undergo proliferation, including stem cells, progenitor cells, or cancer cells. Each cell is either at the resting or proliferative phase during cell cycling. Cells that lose the proliferation ability are removed due to senescence, terminal differentiation, or death. Additionally, the specific cell types are not included in the model, as they are defined with specific patterns of marker gene expressions. The heterogeneity of cells is represented by the variance in the epigenetic states of cells, which are associated with the \textit{kinetotype} of each cell rather than the cell type.

\section{Modeling cellular plasticity}
\label{sec:plast}

The inheritance function $p(\vec{x}, \vec{y})$ is vital to describe the plasticity of cells. However, the exact formula of the inheritance function is challenging to determine biologically, which is dependent on the complex biochemical reactions during the biological process of cell division. Nevertheless, while we consider $p(\vec{x}, \vec{y})$ as a conditional probability density, we can focus on the epigenetic state before and after cell division and omit the intermediate complex process.

Typically, our focus lies on the gene expressions that play an essential role in the gene network. The temporal dynamics of gene expression within one cell cycle can be described by a chemical rate equation of form \eqref{eq:1.10}. Cellular plasticity during cell division leads to the discontinuous changes in the cell states $\vec{x}$ and the parameters $\vec{q}$, and the long-term dynamics can be described by a discontinuous dynamical equation across cell division as \eqref{eq:1.11}. In this way, we can obtain phenomenological formulations for the inheritance function through numerical simulation based on gene regulatory networks and the laws of epigenetic state inheritance during cell division \citep{Huang:2017jr,Huang:2019hn,Sahoo:2021gj}. For more discussions, refer to \cite{Jinzhi-book}.

Alternatively, we can assume the phenomenological formulations directly based on experimental observations. Let the epigenetic state $\vec{x} = (x_1, \cdots, x_n)$ represent $n$ independent state variables, and assume that these states vary independently during cell division, we have
$$
p(\vec{x}, \vec{y}) = \prod_{i=1}^n p_i(x_i, \vec{y}),
$$
where $p_i(x_i, \vec{y})$ means the inheritance function of $x_i$, given the state $\vec{y}$ of the mother cells.

The inheritance function $p_i(x_i, \vec{y})$ can be represented as the density function of $x_i$, which is often phenomenological assumed based on the biological implications. For example, we can take the beta distribution for the normalized nucleosome modifications \citep{Huang:2017jr} or gamma distribution for transcription levels \citep{Cai:2006cp}.

Here, we assume that $0<x_i<1$, and  $p_i(x_i, \vec{y})$ is given by the density function of beta distribution, \textit{i.e.},
\begin{equation}
\label{eq:betadist}
p_i(x_i, \vec{y}) = \dfrac{x_i^{a_i(\vec{y}) - 1} (1-x_i)^{b_i(\vec{y}) - 1}}{B(a_i(\vec{y}), b_i(\vec{y}))},\quad B(a, b) = \dfrac{\Gamma(a) \Gamma(b)}{\Gamma(a + b)},
\end{equation}
where $\Gamma(\cdot)$ represents the gamma function. Here, the inheritance function depends on two shape parameters, $a$ and $b$, which are functions of the epigenetic state $\vec{y}$ of the mother cell. To determine the functions $a_i(\vec{y})$ and $b_i(\vec{y})$ from experimental data, we write the mean and variance of $x_i$, given the state $\vec{y}$, as
\begin{equation}
\label{eq:phi1}
\mathrm{E}(x_i | \vec{y}) = \phi_i(\vec{y}), \quad \mathrm{Var}(x_i | \vec{y}) = \dfrac{1}{1 + \eta_i(\vec{y})} \phi_i(\vec{y}) (1 - \phi_i(\vec{y})).
\end{equation}
Accordingly, the shape parameters can be given by 
\begin{equation}
\label{eq:ab}
a_i(\vec{y}) = \eta_i(\vec{y})\phi_i(\vec{y}), \quad b_i(\vec{y}) = \eta_i(\vec{y}) (1 - \phi_i(\vec{y}))
\end{equation}
through predefined functions $\phi_i(\vec{y})$ and $\eta_i(\vec{y})$. Here, the functions $\phi_i(\vec{y})$ and $\eta_i(\vec{y})$ should always satisfy
$$0 < \phi_i(\vec{y}) < 1, \quad \eta_i(\vec{y}) > 0.$$

The predefined functions, $\phi_i(\vec{y})$ and $\eta_i(\vec{y})$, are used to determine the conditional expectation and variance of the experimental data at the single cell level, as seen in equation \eqref{eq:phi1}. These functions act as a link between the model formulation and the experimental data. Additionally, $\phi_i(\vec{y})$ defines how the epigenetic state, $x_i$, depends on other components. This is often linked to the gene regulation network that underlies epigenetic states, and hence $\phi_i(\vec{y})$ can incorporate the information about gene networks.

A simple example of the above formulations can be shown by a single epigenetic state $x\in [0, 1]$, representing the stemness of a cell. Based on biological implications, cells with high stemness exhibit a low proliferation rate and an extremely low differentiation rate, the cells with intermediate stemness have a high proliferation rate and low differentiation rate, and the cells with low stemness have the lowest proliferation rate and the highest differentiation rate. Thus, we can formulate the proliferation rate $\beta$ and the differentiation rate $\kappa$ as (refer to \cite{Lei:2020fw,Zhang:2021fs})
\begin{equation}
\label{eq:bck}
\beta(c, x) = \bar{\beta} \times \dfrac{\theta^{m_0}}{\theta^{m_0} + c^{m_0}} \times \dfrac{a_1 x + (a_2 x)^2}{1 + (a_3 x)^6},\quad \kappa(x) = \kappa_0 \times \dfrac{1}{1+ (b_1x)^6},
\end{equation}
where
$$
c(t) = \int_0^1 Q(t, x) \zeta(x) \mathrm{d} x.
$$
Here, we refer to equation \eqref{eq:beta} to define the proliferation rate $\beta$ and assume $\beta_1(x) \equiv 0$. We should note that the mathematical form \eqref{eq:bck} is not unique. Any form describing the qualitative dependencies of $\beta$ and $\kappa$ on the stemness $x$ would be acceptable. 

Moreover, we assume that the apoptosis rate $\mu$ and the proliferating phase duration $\tau$ are independent of the stemness $x$. 

The equation for heterogeneous stem cell regeneration becomes
\begin{equation}
\label{eq:1x}
\begin{array}{rcl}
\displaystyle \dfrac{\partial Q(t, x)}{\partial t} &=& \displaystyle -Q(t, x) (\beta(c(t), x) + \kappa(x)) \\
&&\displaystyle {}+ 2 e^{-\mu \tau}\int_0^1 \beta(c(t - \tau), y) Q(t - \tau, y) p(x, y) \mathrm{d} y,
\end{array}
\end{equation}
where the inheritance function
$$
p(x, y) = \dfrac{x^{a(y) - 1} (1 - x)^{b(y) - 1}}{B(a(y), b(y))}, \quad B(a, b) = \dfrac{\Gamma(a) \Gamma(b)}{\Gamma(a + b)}.
$$ 

Equation \eqref{eq:1x} provides a simple example for mathematical studies. However, many fundamental problems remain open, such as the existence and uniqueness of the steady-state solution, the stability of steady-state solutions, and the existence of oscillatory solutions. For discussions of these problems, please refer to \cite{Lei:2020ei} or Section \ref{sec:mp0} in this chapter. 

\section{Individual-based model of multicellular tissues}
Equation \eqref{eq:6.26} provides a general mathematical framework to model stem cell regeneration when heterogeneity and plasticity of epigenetic or genetic states are included. This framework can describe many biological processes associated with stem cell regeneration, including development, aging, and cancer evolution \citep{Lei:2020fw}. Nevertheless, solving the equation numerically \eqref{eq:6.26} is expensive when high-dimensional epigenetic states are considered. Based on the above framework, we often develop hybrid computational models for multicellular tissues in applications.

Based on the mathematical framework \eqref{eq:6.26}, a hybrid numerical scheme was developed that combines a discrete stochastic process for the epigenetic/genetic states of individual cells with a continuous model of cell population growth \citep{Lei:2020fw}. In numerical simulation, a multicellular system is represented by a collection of multiple cells $C_i$, each cell has its epigenetic states $\vec{x}_i$, \textit{i.e.}, the system $\Omega_t = \left\{[C_i(\vec{x}_i)]_{i=1}^{Q(t)}\right\}$, where $Q(t)$ represents the number of resting phase stem cells at time $t$. During a time interval $(t, t + \Delta t)$, each cell ($C_i(\vec{x}_i)$) undergoes cell fate decision, \textit{e.g.}, proliferation, apoptosis, or terminal differentiation, with a probability given by the kinetic rates. The probabilities of proliferation, apoptosis, or differentiation are given by $\beta(c, \vec{x}_i) \Delta t, \mu(\vec{x}_i) \Delta t$, or $\kappa(\vec{x}_i) \Delta t$, respectively, and hence are dependent on the epigenetic state of each cell as well as the microenvironmental condition $c = \int Q(t, \vec{x}) \zeta(\vec{x}) \mathrm{d} \vec{x}$. The total cell number $Q(t)$ changes after a time step $\Delta t$ following the behaviors of all cells. When a cell undergoes proliferation, the epigenetic states of daughter cells change randomly according to the transition function $p(\vec{x}, \vec{y})$. In this hybrid model, all detailed molecular interactions are hidden within the kinetic rates and the inheritance function. Single-cell-based models can implement the proposed hybrid model through GPU architecture \citep{Song:2018kl}.

The above hybrid numerical scheme can also be integrated with the stochastic modeling of gene networks, through which the gene expression dynamics in individual cells are described with mathematical models of the form \eqref{eq:1.17} between cell divisions. Each cell's kinetotype ($\beta$, $\mu$, $\kappa$, and $\tau$) depends on the gene expression state $\vec{x}$. For a detailed numerical scheme, refer to \cite{Huang:2023hi}. For more examples, please refer to \cite{Lei:2020fw,Zhang:2021fs,Liang:2023aa,Huang:2023hi,Zhang:2022vd}.

\section{Waddington landscape}
\label{sec:wad}

Understanding cell fate decisions and cell differentiation is crucial in biology. Waddington's epigenetic landscape is a fundamental concept in this understanding \cite{Waddington:2012bu}. The concept visualizes a cell as a ball rolling on a mountain, where valleys correspond to stable cell phenotypes and ridges represent cell fate choices leading to new phenotypes. In other words, the landscape analogy illustrates how a cell's gene expression and environment interact to determine its development into a specific cell type. Mathematically, the definition of the Waddington landscape and its evolution are crucial in telling us more than reasoning about the process of cell fate decision during tissue development. 

The Waddington landscape refers to the probability of a cell choosing a specific phenotype while developing. Thus, we can define the Waddington landscape as akin to the energy landscape in descriptions of physical systems, and the cell state probability is analogous to the Boltzmann distribution in statistical mechanics. Thus, we define the Waddington landscape as a potential $U(\vec{x})$, where $\vec{x}$ represents the state or phenotype of a cell. The likelihood of a cell in a state $\vec{x}$ is given by  
$$
P(\vec{x})  = C e^{- \gamma U(\vec{x})}
$$
according to the Boltzmann distribution, or the Arrhenius equation\cite{Hanggi:1990ud,Zhou:2012kf}, where $\gamma$ is a constant, and $C$ is a normalization coefficient. Conversely, the Waddington landscape is associated with the probability density $P(\vec{x})$ through (up to a scaling factor and constant translation)
\begin{equation}
U(\vec{x}) = - \log P(\vec{x}).
\end{equation}
The potential $U(\vec{x})$ usually depends on the microenvironment conditions during tissue development and evolves with time. Hence, the evolutions of the population density and the Waddington landscape are represented by the time-dependent functions $P(t, \vec{x}$) and $U(t, \vec{x})$, respectively. 

\subsection{Gene circuit dynamics approach}

There are two methods to formulate the temporal evolution of the potential $U(t, \vec{x})$.  The gene circuit dynamics \eqref{eq:1.8} under a noisy fluctuating environment can be formulated as
\begin{equation}
\label{eq:sde1}
\dfrac{\mathrm{d} \vec{x}}{\mathrm{d} t} = \vec{F}(\vec{x}) + \vec{\eta},
\end{equation}   
where $\vec{\eta} = (\eta_1, \eta_2, \cdots, \eta_n)$ is a multi-dimensional Gaussian noise term. The correlation satisfies $\langle \eta_i(\vec{x}, t_1) \eta_j(\vec{x}, t_2) \rangle = 2 D \delta_{i,j}\delta(t_1 -  t_2)$, with $D$ as the diffusion coefficient. Considering a system of a multi-cellular system where the gene expressions of all cells follow an initial distribution density $P(0, \vec{x})$, the evolution of the probability density $P(t, \vec{x})$ can be formulated through the Fokker-Planck equation
\begin{equation}
\label{eq:fp1}
\dfrac{\partial P(t, \vec{x})}{\partial t} + \nabla \cdot \vec{J}(t, \vec{x}) = 0,
\end{equation}
where $\nabla = (\frac{\partial\ }{\partial x_1},\cdots, \frac{\partial\ }{\partial x_n})$, and $\vec{J}(t, \vec{x})$ is the probability flux defined as
\begin{equation}
\label{eq:J0}
\vec{J}  = \vec{F} P - D \nabla P.
\end{equation}
The probability flux consists of two components: the drift $\vec{F} P$, which quantifies the inclination of the cell state to move in a specific direction, and the diffusion term $D \nabla P$, accounting for the random fluctuation in the cell state. Moreover, $U(t, \vec{x}) = - \log P(t, \vec{x})$ gives the evolution of the Waddington landscape. 

From the equation \eqref{eq:fp1}, at the stationary state, the divergence of the probability flux $\vec{J}_{ss}$ vanishes $\nabla\cdot \vec{J}_{ss} = 0$, yielding the steady-state probability density $P_{ss}(\vec{x})$. We note $P_{ss}(\vec{x}) = e^{-U_{ss}(\vec{x})}$ at the steady-state, where $U_{ss}(\vec{x})$ represents the steady-state potential. Thus, equation \eqref{eq:J0} results in a decomposition of the force \cite{Lapidus:2008iq,Wang:2011hz}
\begin{equation}
\label{eq:df}
\vec{F}(\vec{x}) = - D \nabla U_{ss}(\vec{x}) + \vec{J}_{ss}(\vec{x})/P_{ss}(\vec{x}).
\end{equation} 
Equation \eqref{eq:df} decomposes the force term of the gene circuit dynamics into two components. The first part is the potential gradient, connected to the steady-state probability by $U = - \ln P_{ss}$. The second part is the curl flux force, establishing a link between the divergence-free steady-state probability flux $\vec{J}_{ss}$ and the steady-state probability $P_{ss}$. 

Determining the landscape $U$ involves obtaining the steady-state solution of the Fokker-Planck equation \eqref{eq:fp1}. This equation describes changes in the probability density of the molecular states of a cell. However, it may not be suitable for describing tissue development because it does not account for biological processes such as cell division and death. 
 
To incorporate the process of cell division and cell death, we replace $P(t,\vec{x})$ with the cell population density $f(t, \vec{x})$ and introduce a coefficient $R(\vec{x})$ to represent the birth-day rate (BDR) of a cell with state $\vec{x}$. Cell proliferation occurs when $R(\vec{x}) > 0$, while cell death occurs when $R(\vec{x})< 0$. Following the discussion proposed by \cite{Shi:2022gh}, consider a cell $\omega$ with gene expression $\vec{X}_t(\omega)$ starting from $\vec{Y}_0$ at $t=0$. The corresponding weighted stochastic dynamics in the It\^{o} sense can be written as
\begin{equation}
\label{eq:37}
\left\{
\begin{aligned}
&\mathrm{d} \vec{X}_t(\omega) =  \vec{F}(\vec{X}_t(\omega)) \mathrm{d} t + \sqrt{2 D}\, \mathrm{d} \vec{W}_t(\omega),\\
&\vec{X}_t(\omega) |_{t = 0}  = \vec{Y}_0(\omega),\\
& \mathrm{d} \rho_t(\omega) = R(\vec{X}_t(\omega)) \rho_t(\omega) \mathrm{d} t,\\
&\rho_t(\omega)|_{t=0} = 1.
\end{aligned}
\right.
\end{equation}
Here $\rho_t(\omega)$ is a time-varying weight for cell $\omega$. The population density $f(t, \vec{x})$ is linked to the above equation as
\begin{equation}
\label{eq:38}
f(t, \vec{x}) = \mathbb{E}\{\rho_t(\omega) \delta( \vec{x} - \vec{X}_t(\omega))\},
\end{equation}
where $\delta$ is the Dirac delta function, the expectation is taken over all possible trajectories $\omega$. 

The equations \eqref{eq:37} and \eqref{eq:38} result in the following population balance equation (refer to \cite{Shi:2022gh,Weinreb:2018dg} for detailed discussions)
\begin{equation}
\label{eq:cf}
\dfrac{\partial f}{\partial t} = \nabla \cdot (D \nabla f) - \nabla \cdot (f \vec{F}) + R f.
\end{equation} 
Given the BDR $R(\vec{x})$, equation \eqref{eq:cf} describes the time evolution of the population density. From \eqref{eq:cf}, the following constraint
$$
\int R(\vec{x}) f_{ss}(\vec{x}) \mathrm{d} \vec{x} = 0
$$
for $R(\vec{x})$ is required to ensure a biologically meaningful steady-state population density $f_{ss}(\vec{x})$.

As discussed in \cite{Shi:2022gh}, denote by $P_U(x)$ the steady-state population density with the known  BDR $R(\vec{x})$ and by $P_0(\vec{x})$ the steady-state population density with $R(\vec{x}) \equiv 0$ for \eqref{eq:cf}. Then, two energy landscapes can be constructed as
\begin{equation}
U_{ss}(\vec{x}) = -  \log P_U(\vec{x})
\end{equation}  
and
\begin{equation}
V_{ss}(\vec{x}) = -  \log P_0(\vec{x})/P_U(\vec{x}).
\end{equation}
The potential $U(\vec{x})$ drives the system or cells to the steady distribution whose metastable basins indicate cell types. While $V(\vec{x})$ quantifies the changes in the potential caused by the influence of cell proliferation and death, the values of $V$ depict the pluripotency, and its negative gradient field describes the differentiation direction. Thus, we can consider $U_{ss}(\vec{x}) + V_{ss}(\vec{x})$ as the Waddington landscape from a stem cell state to a differentiated cell state \cite{Shi:2022gh}.

Through the two potentials $U_{ss}(\vec{x})$ and $V_{ss}(\vec{x})$, we can decompose the force terms as
\begin{equation}
\label{eq:df1}
F(\vec{x}) = - D \nabla U_{ss}(\vec{x}) - D \nabla V_{ss}(\vec{x}) + \vec{G}(\vec{x}),
\end{equation}
where $\vec{G}(\vec{x})$ is the curl part that describes the non-gradient nature of the considered dynamics.

Moreover, substituting the force decomposition \eqref{eq:df1} into \eqref{eq:cf}, we have
\begin{eqnarray*}
\dfrac{\partial f}{\partial t} &=& \nabla \cdot (D \nabla f) - \nabla \cdot (f (- D \nabla (U_{ss}(\vec{x}) + V_{ss}(\vec{x})) + G(\vec{x}))) + R(\vec{x}) f\\
&=& \nabla \cdot (D (\nabla f  + f \nabla (U_{ss}(\vec{x}) + V_{ss}(\vec{x})))) - \nabla \cdot (f G(\vec{x})) + R(\vec{x}) f.
\end{eqnarray*}
Thus, we obtain the population balance equation of the population density $f(t, \vec{x})$:
\begin{equation}
\label{eq:34}
\dfrac{\partial f}{\partial t} = \nabla \cdot (D (\nabla f + f \nabla (U_{ss}(\vec{x}) + V_{ss}(\vec{x})))) - \nabla \cdot(f \vec{G}(\vec{x})) + R(\vec{x}) f.
\end{equation}
The equation \eqref{eq:34} establishes the connection between population density and the stationary state probability density and the flux. Through the general framework equations \eqref{eq:cf} or \eqref{eq:34}, various data-based methods have been proposed to identify cluster/cell types, differentiation trajectories, pseudotime, and cell pluripotency from the experimental data, such as the population balance analysis (PBA) \cite{Weinreb:2018dg,Briggs:2018aa,Fischer:2019cia} and landscape of differentiation dynamics (LDD) \cite{Shi:2022gh,Shi:2019gd}.  

Finally, it is important to point out that there are different ways to construct the potential landscape from a gene circuit dynamics of the form \eqref{eq:sde1}. Here, we mainly introduced Wang's potential landscape defined from the steady-state distribution associated with the Fokker-Planck equation \eqref{eq:fp1} \cite{Lapidus:2008iq,Wang:2011hz}. Alternative definitions include the Freidlin-Wentzell quasi-potential from the large deviation theory \cite{Freidlin:1984kd}, and Ao's potential through stochastic differentiation decomposition and a generalized Lyapunov function \cite{Ao:2004gd,Wang:2023gd}. A summary of different types of potential can be found in \cite{Zhou:2016gl}.

\subsection{Stem cell regeneration approach}
\label{sec:scra}

The equation \eqref{eq:cf} mentioned above is derived from the Fokker-Planck approach, following the stochastic differentiation equation \eqref{eq:sde1} for gene circuit dynamics. However, these equations do not account for cell division and plasticity. On the other hand, we can use equation \eqref{eq:6.26}, which explicitly includes cell division and plasticity, to derive the evolution dynamics of cell population density for stem cell regeneration.

We note that 
\begin{equation}
Q(t) = \int_\Omega Q(t,\vec{x}) \mathrm{d} \vec{x}
\end{equation}
represents the total cell number. The relative cell number with epigenetic state $\vec{x}$ is given by
\begin{equation}
\label{eq:fd}
f(t, \vec{x}) = \dfrac{Q(t, \vec{x})}{Q(t)}.
\end{equation}
It is easy to have
\begin{equation}
\label{eq:fint}
\int_\Omega f(t, \vec{x}) \mathrm{d} \vec{x} \equiv 1,\forall t>0.
\end{equation}
The function $f(t, \vec{x})$ depicts the evolution of the probability density of epigenetic states, and the Waddington landscape is given by
\begin{equation}
\label{eq:wd}
U(t, \vec{x}) = -\log f(t, \vec{x}).
\end{equation}

From equations \eqref{eq:6.26}, \eqref{eq:sfg}, \eqref{eq:fd}, and \eqref{eq:fint}, we can derive the evolution equation for $f(t, \vec{x})$ as following. 
\begin{eqnarray*}
\dfrac{\partial f(t, \vec{x})}{\partial t} &=& -\dfrac{Q(t, \vec{x})}{Q(t)^2} \dfrac{\mathrm{d} Q}{\mathrm{d} t} + \dfrac{1}{Q(t)} \dfrac{\partial Q(t, \vec{x})}{\partial t}\\
&=& - f(t, \vec{x}) \dfrac{1}{Q(t)} \Big(-\int_\Omega Q(t, \vec{y}) (\beta(c, \vec{y}) + \kappa(\vec{y})) \mathrm{d} \vec{y}\\
&&{}\qquad\qquad\qquad + 2 \int_\Omega \beta(c_{\tau(\vec{y})},\vec{y}) Q(t - \tau(\vec{y}), \vec{y}) e^{-\mu(\vec{y})\tau(\vec{y})} \mathrm{d} \vec{y}\Big)\\
&&{} + \dfrac{1}{Q(t)}\Big(-Q(t, \vec{x}) (\beta(c,\vec{x})  + \kappa(\vec{x}))\\
&&{}\qquad\qquad + 2 \int_{\Omega} \beta(c_{\tau(\vec{y})},\vec{y}) Q(t-\tau(\vec{y}),\vec{y}) e^{-\mu(\vec{y})\tau(\vec{y})} p(\vec{x},\vec{y}) \mathrm{d}\vec{y}\Big)\\
&=&f(t,\vec{x}) \int_\Omega f(t,\vec{y})(\beta(c, \vec{y}) + \kappa(\vec{y}))\mathrm{d} \vec{y} - f(t, \vec{x}) (\beta(c, \vec{x}) + \kappa(\vec{x}))\\
&&{} - \dfrac{2}{Q(t)}\int_\Omega \beta(c_{\tau(\vec{y})}, \vec{y}) Q(t - \tau(\vec{y}), \vec{y}) e^{-\mu(\vec{y}) \tau(\vec{y})} f(t,\vec{x}) \mathrm{d} \vec{y}\\
&&{} + \dfrac{2}{Q(t)}\int_\Omega \beta(c_{\tau(\vec{y})}, \vec{y}) Q(t - \tau(\vec{y}), \vec{y}) e^{-\mu(\vec{y}) \tau(\vec{y})} p(\vec{x}, \vec{y}) \mathrm{d} \vec{y}\\
&=&\dfrac{2}{Q(t)}\int_\Omega \beta(c_{\tau(\vec{y})}, \vec{y}) Q(t - \tau(\vec{y}), \vec{y}) e^{-\mu(\vec{y}) \tau(\vec{y})}(p(\vec{x}, \vec{y}) - f(t, \vec{x}))\mathrm{d} \vec{y}\\
&&{} - f(t, \vec{x}) \int_\Omega f(t, \vec{y}) ((\beta(c, \vec{x}) + \kappa(\vec{x})) - (\beta(c, \vec{y}) + \kappa(\vec{y})))\mathrm{d} \vec{y}.
\end{eqnarray*}
Thus, we have the equation
\begin{equation}
\label{eq:fvh}
\begin{aligned}
\dfrac{\partial f(t, \vec{x})}{\partial t} =&
\dfrac{2}{Q(t)} \int_\Omega \beta(c_{\tau(\vec{y})}, \vec{y}) Q(t - \tau(\vec{y}), \vec{y}) e^{-\mu(\vec{y}) \tau(\vec{y})} (p(\vec{x}, \vec{y}) - f(t, \vec{x})) \mathrm{d} \vec{y} \\
&{}-f(t, \vec{x}) \int_\Omega f(t, \vec{y})\left((\beta(c, \vec{x}) + \kappa(\vec{x})) - (\beta(c, \vec{y}) + \kappa(\vec{y}))\right) \mathrm{d} \vec{y}.
\end{aligned}
\end{equation}
Moreover, since
$$
\dfrac{\partial U(t, \vec{x})}{\partial t} = -\frac{1}{f(t, \vec{x})} \dfrac{\partial f(t, \vec{x})}{\partial t},
$$
we have
\begin{eqnarray}
\label{eq:Wdf}
\dfrac{\partial U(t, \vec{x})}{\partial t} &=& -
\dfrac{2}{Q(t,\vec{x})} \int_\Omega \beta(c_{\tau(\vec{y})}, \vec{y}) Q(t - \tau(\vec{y}), \vec{y}) e^{-\mu(\vec{y}) \tau(\vec{y})} (p(\vec{x}, \vec{y}) - f(t, \vec{x})) \mathrm{d} \vec{y} \nonumber\\
&&{}+ \int_\Omega f(t, \vec{y}) \left((\beta(c, \vec{x}) + \kappa(\vec{x})) - (\beta(c, \vec{y}) + \kappa(\vec{y}))\right) \mathrm{d} \vec{y}.
\end{eqnarray}
The equation \eqref{eq:Wdf} gives the evolution of the Waddington landscape of a system of stem cell regeneration with cell heterogeneity and plasticity.

When the system reaches the equilibrium state so that $Q(t)$ and $f(t, \vec{x})$ are independent of the time $t$,  we write
$$
Q(t) = Q^*, f(t, \vec{x}) = f^*(\vec{x}), U(t, \vec{x}) = U^*(\vec{x}), c(t) = \int Q(t, \vec{x}) \zeta(\vec{x}) \mathrm{d}\vec{x} = c^*.
$$
The equation \eqref{eq:fvh} becomes
\begin{equation}
\label{eq:ghd}
\begin{aligned}
&2 \int_\Omega \beta(c^*, \vec{y}) e^{-\mu(\vec{y}) \tau(\vec{y})} f^*(\vec{y})(p(\vec{x}, \vec{y}) - f^*(\vec{x})) \mathrm{d} \vec{y}\\
&-f^*(\vec{x}) \int_\Omega f^*(\vec{y})\left((\beta(c^*, \vec{x}) + \kappa(\vec{x})) - (\beta(c^*, \vec{y}) + \kappa(\vec{y}))\right) \mathrm{d} \vec{y} = 0
\end{aligned}
\end{equation}
at the equilibrium state. This gives an integral equation for the population density at the equilibrium state. Define a nonlinear operator $\mathcal{F}_c$ as
$$
\begin{aligned}
\mathcal{F}_c[f] =
&2 \int_\Omega \beta(c, \vec{y}) e^{-\mu(\vec{y}) \tau(\vec{y})} f(\vec{y})(p(\vec{x}, \vec{y}) - f(\vec{x})) \mathrm{d} \vec{y} \nonumber\\
&-f(\vec{x}) \int_\Omega f(\vec{y})\left((\beta(c, \vec{x}) + \kappa(\vec{x})) - (\beta(c, \vec{y}) + \kappa(\vec{y}))\right) \mathrm{d} \vec{y},
\end{aligned}
$$
the equation \eqref{eq:ghd} gives a nonlinear eigenvalue problem
\begin{equation}
\mathcal{F}_{c}[f] = 0.
\end{equation}
The equilibrium state density function $f^*(\vec{x})$ corresponds to the eigenfunction of the operator $F_c$ with a positive eigenvalue $c^*$. Accordingly, the Waddington landscape at the equilibrium state is 
\begin{equation}
U^*(\vec{x}) = - \log f^*(\vec{x}).
\end{equation}
Thus, the mathematical formulation provides a general method of calculating the evolution of Waddington's landscape during tissue growth.

\subsection{Combining the gene circuit dynamics with stem cell regeneration}
\label{sec:comb}

We have introduced two methods to formulate the population density $f(t, \vec{x})$. The gene circuit dynamics approach results in equation \eqref{eq:cf}, where a birth-death rate $R(\vec{x})$ is introduced to account for the effect of cell birth and death. The stem cell generation approach considers the heterogeneity and plasticity of cells during cell divisions and explicitly formulates the biological processes of proliferation, differentiation, and apoptosis through the kinetotype of cells. These two approaches separately describe the mechanisms of cell type switches, driven by noise perturbations to the gene network or epigenetic changes during cell division. Here, we propose an integrated approach that combines both processes.

In equation \eqref{eq:cf}, the birth-death rate $R(\vec{x})$ depends solely on the state of the cell, excluding cell-to-cell interactions. Additionally, cell plasticity associated with cell division is not considered in \eqref{eq:cf}. However, the stem cell regeneration approach, as indicated by equation \eqref{eq:fvh}, suggests a growth operator $\mathcal{R}$ for the population density as follows:
\begin{equation}
\begin{aligned}
\mathcal{R}[f] = &\dfrac{2}{Q(t)} \int_\Omega \beta(c_{\tau(\vec{y})}, \vec{y}) Q(t - \tau(\vec{y}), \vec{y}) e^{-\mu(\vec{y}) \tau(\vec{y})} (p(\vec{x}, \vec{y}) - f(t, \vec{x})) \mathrm{d} \vec{y} \\
&{}-f(t, \vec{x}) \int_\Omega f(t, \vec{y})\left((\beta(c, \vec{x}) + \kappa(\vec{x})) - (\beta(c, \vec{y}) + \kappa(\vec{y}))\right) \mathrm{d} \vec{y}.
\end{aligned}
\end{equation}
This operator takes into account the regulation of cell proliferation, differentiation, and apoptosis through both microenvironment conditions (via the factor $c$) and cellular states (via the epigenetic state $\vec{x}$). Moreover, cell plasticity during cell division is also involved through the inheritance function $p(\vec{x}, \vec{y})$.

Replacing the birth-death term $R f$ with the growth operator $\mathcal{R}[f]$, the population balance equation \eqref{eq:34} can be expressed as
\begin{equation}
\label{eq:45}
\dfrac{\partial f}{\partial t} =\nabla \cdot (D\nabla f) -\nabla\cdot (f \vec{F}) + \mathcal{R}[f].
\end{equation}
Thus, the equation \eqref{eq:45} together with \eqref{eq:6.26}-\eqref{eq:sfg}, provides an integrative mathematical model for the evolution of the population density $f(t, \vec{x})$. This equation combines the dynamics of the gene regulation network with heterogeneous stem cell regeneration. Accordingly, $U(t, \vec{x}) = - \log f(t, \vec{x})$ gives the evolution of Waddington's epigenetic landscape.

In particular, we consider a simple situation where the heterogeneities in the kinetotype of cells are omitted, \textit{i.e.}, the rate functions $\beta, \kappa, \mu, \tau$ are independent of the epigenetic state $\vec{x}$. Moreover, we assumed that total cell number $Q(t)$ approaches a stable state $Q(t) = Q^*$, and the factor $c = \int Q(t, \vec{x}) \mathrm{d} \vec{x} = Q^*$. Thus, the growth factor becomes
$$
\mathcal{R}[f] = 2 \beta^* e^{-\mu \tau}  \left( \int  f(t-\tau, \vec{y}) p(\vec{x}, \vec{y}) \mathrm{d}\vec{y} - f(t, \vec{x})\right).
$$
Here, $\beta^*=\beta(Q^*)$ represents the proliferation rate at the steady state given by \eqref{eq:bs}. Thus, the equation \eqref{eq:45} becomes
\begin{equation}
\label{eq:47}
\dfrac{\partial f}{\partial t} =\nabla \cdot (D\nabla f) -\nabla\cdot (f \vec{F}) + 2 \beta^* e^{-\mu \tau}  \left( \int  f(t-\tau, \vec{y}) p(\vec{x}, \vec{y}) \mathrm{d}\vec{y} - f(t, \vec{x})\right).
\end{equation}
At the stationary state, the stationary densify $f^*(\vec{x})$ satisfies a nonlocal elliptic equation
\begin{equation}
\label{eq:48}
\nabla \cdot (D\nabla f^*) -\nabla\cdot (f^* \vec{F}) = - 2 \beta^* e^{-\mu \tau}  \left( \int  f^*(\vec{y}) p(\vec{x}, \vec{y}) \mathrm{d}\vec{y} - f^*(\vec{x})\right).
\end{equation}
The mathematical formulations of the Waddington landscape of stem cell regeneration with cell plasticity are given by equations \eqref{eq:47} and \eqref{eq:48}. Many basic mathematical problems associated with these equations still need to be solved.

\section{Applications}

The mathematical framework for stem cell regeneration presented in this chapter is versatile, allowing for its application in modeling the dynamics of multi-cellular system development, encompassing processes such as tissue development and tumor progression. Here, we introduce three instances where this modeling framework has been applied to investigate tumor progression and cell type transitions. For in-depth exploration and comprehensive details, please refer to the associated references.

\subsection{Abnormal growth induced by changes in the microenvironmental conditions}

We applied the model framework to describe the process of abnormal growth of tumor cells induced by changes in the microenvironmental conditions (refer to  \cite{Zhang:2022vd}). Consider a system of cells where the epigenetic state of each cell is denoted as $\vec{x} = (x_1, x_2)$, with $x_1$ representing the stemness of a cell and $x_2$ indicating the malignancy of a cell. Both $x_1$ and $x_2$ are normalized to the interval $[0,1]$, defining the epigenetic state $\vec{x}\in \Omega = [0,1]\times [0, 1]$.

Referring to the expressions in \eqref{eq:bck}, we assume that the proliferation rate $\beta$ and differential rate $\kappa$ depend on the stemness $x_1$, and are defined as
\begin{equation}
\beta(c, \vec{x}) = \beta_0(\vec{x}) \dfrac{1}{1 + (c/\theta)^n} ,\quad \beta_0(\vec{x}) = \beta_0 \dfrac{a_1 x_1 + (a_2 x_1)^{s_1}}{1 + (a_3 x_1)^{s_1}}
\end{equation}
and
\begin{equation}
\kappa(\vec{x}) = \dfrac{\kappa_0}{1 + (b_1 x_1)^{s_1}}.
\end{equation}
Here, we take $c$ as the cell number $Q$. The coefficient $\theta(\vec{x})$ represents the repression of cell proliferation through cell responses to micro-environmental cytokines and is dependent on malignancy. Thus, we assume that $\theta$ increases with the malignancy $x_2$, leading to 
\begin{equation}
\theta(\vec{x}) = \theta_0 + \theta_1 \dfrac{x_2^{s_2}}{\theta_2^{s_2} + x_2^{s_2}},
\end{equation}
where $\theta_0$, $\theta_1$, and $\theta_2$ are parameters.

The microenvironmental condition may affect the fitness of a cell in a given environment. To consider this effect on cancer evolution, we introduced a microenvironment index $u$, representing the effects of the microenvironment on malignancy and cell survival. Assuming $0<u<1$, with larger $u$ indicating a microenvironment more suitable for cells with higher malignancy, we can define the fitness of a cell as
$$
g(u, x_2) = g_0 x_2^u (1-x_2)^{1-u},
$$
where $g_0$ is a constant. Moreover, we assume that the apoptosis rate $\mu$ is dependent on the microenvironmental index $u$ and malignancy $x_2$, expressed as 
\begin{equation}
\label{eq:mu0}
\mu(u, x_2) = \dfrac{\mu_0}{1 + \rho e^{g(u, x_2)}}.
\end{equation}
Here, $\mu_0$ and $\rho$ are constants, indicating a maximum apoptosis rate $\mu_0/(1+\rho)$ when the fitness $g  = 0$. Better fitness implies a lower apoptosis rate of the cell. 

Similar to the previous argument, the inheritance function
$$
p(\vec{x}, \vec{y})  = p_1(x_1, \vec{y})\times p(x_2 , \vec{y}),
$$
where $p_i(x_i, \vec{y})$ are density functions of beta distribution
$$
p_i(x_i, \vec{y}) = \dfrac{x_i ^{a_i(\vec{y}) - 1} (1-x_i)^{b_i(\vec{y}) - 1}}{B(a_i(\vec{y}), b_i(\vec{y}))},\quad B(a, b) = \dfrac{\Gamma(a) \Gamma(b)}{\Gamma(a + b)}.
$$
The shape parameters $a_i(\vec{y})$ and $b_i(\vec{y})$ are defined by the predefined functions $\phi_i(\vec{y})$ and $\eta_i(\vec{y})$ according to \eqref{eq:phi1} and \eqref{eq:ab}. Specifically, we take $\eta_1(\vec{y}) = \eta_2(\vec{y}) = \eta$ as constants and let $\phi_1(\vec{y}) = \phi_1(y_1)$ and  $\phi_2(\vec{y}) = \phi_2(y_2)$ be defined as
\begin{eqnarray}
\phi_1(y_1) &=& c_1 + d_1 \times \dfrac{(\alpha_1 y_1)^{1.5}}{1 + (\alpha_1 y_1)^{1.5}},\\
\phi_2(y_2) &=& c_2 + d_2 \times \dfrac{(\alpha_2 y_2)^{2.1}}{1 + (\alpha_2 y_2)^{2.1}}.
\end{eqnarray}
Here, $c_1$, $c_2$, $d_1$ are constants, and $d_2$ may depend on the micro-environmental index $u$.  When the microenvironment becomes abnormal (increases of $u$), the cells tend to be more malignant, so $d_2$ increases with $u$. 

An individual-based modeling based on the above formulations can reveal abnormal cell growth dynamics when the micro-environmental index changes from a normal value ($u = 0.1$) to an abnormal value ($u = 0.9$). For detailed discussions of the simulation results, refer to \cite{Zhang:2022vd}.

\subsection{Cell plasticity induced immune escape after CAR-T therapy}

Cancer immunotherapy has marked a significant breakthrough in recent years. However, immune escape frequently occurs following immunotherapy administration \citep{Kim:2007gj,Horn:2020jf,Hegde:2020ba}. Here, we present an example illustrating how the mathematical framework detailed above can describe cancer cell plasticity-induced immune escape after chimeric antigen receptor (CAR) T cell therapy. For a comprehensive discussion, please refer to \cite{Zhang:2021fs}.

CAR-T therapy targeting CD19 has proven effective against B-cell acute lymphoblastic leukemia (B-ALL). While many patients achieve complete response with a single infusion of CD19-targeted CAR-T cells, many experience relapse after therapy \cite{Bhoj:2016aa,Schuster:2017aa}. Our recent experiment showed that relapsed tumors in mice after infusion with CD19-28z-T cells maintain CD19 expression but exhibit a subpopulation of $\mathrm{CD19}^+\mathrm{CD34}^+$  and $\mathrm{CD123}^+\mathrm{CD34}^+$ tumor cells, absent in control NGFT-28z-treated mice \cite{Zhang:2021fs}. Based on this observation, we proposed key assumptions that CAR-T-induced tumor cells transition into hematopoietic stem-like cells (by promoting CD34 expression) and myeloid-like cells (by promoting CD123 expression), thereby evading CAR-T cell targeting \cite{Zhang:2021fs}. 

According to these assumptions, each cell's epigenetic state is represented by marker genes CD19, CD22, CD34, and CD123, pivotal in CD19 CAR-T cell responses and cell lineage dynamics. The proliferation rate $\beta$ and differentiation rate $\kappa$ depend on CD34 expression, a marker of stemness, as follows:
\begin{eqnarray*}
\beta &=&\beta_0 \dfrac{\theta}{\theta + N} \times \dfrac{5.8 [\mathrm{CD34}] + (2.2 [\mathrm{CD34}])^6}{1 + (3.75 [\mathrm{CD34}])^6},\\
\kappa &=& \kappa_0 \dfrac{1}{1 + (4.0 [\mathrm{CD34}])^6}.
\end{eqnarray*}
Here, $N$ represents the total cell number. The apoptosis rate $\mu$ comprises a basal rate $\mu_0$ and a rate associated with the CAR-T signal:
$$
\mu = \mu_0 + \mu_1 \times \mathrm{Signal},
$$
where $\mathrm{Signal}$ represents the CAR-T signal and is defined as
$$
\begin{aligned}
&\mathrm{Signal} = f([\mathrm{CD34}], [\mathrm{CD123}]) \dfrac{\gamma_{19}[\mathrm{CD19}]}{1 + \gamma_{19}[\mathrm{CD19}]+\gamma_{22}[\mathrm{CD22}]}R(t),\\
&f([\mathrm{CD34}], [\mathrm{CD123}]) = \dfrac{1}{(1 + ([\mathrm{CD34}]/X_0)^{n_0}) (1 + ([\mathrm{CD123}]/X_1)^{n_1})}.
\end{aligned}
$$
Here, $R(t)$ is the predefined CAR-T activity. CD34 and CD123, markers of stem-like and myeloid-like cells, respectively, were assumed to inhibit CAR-T signaling.

Cell plasticity is defined similarly to the previous discussions. For instance, given the CD34 expression level in cycle $k$ (denoted by $u_k$), the expression level for cycle $k+1$ (represented by $u_{k+1}$) is a random number from a beta distribution with a probability density
$$
P(u_{k+1} = u | u_k) = \dfrac{u^{a_{34}-1} (1-u)^{b_{34}-1}}{B(a, b)},\quad B(a, b) = \dfrac{\Gamma(a) \Gamma(b)}{\Gamma(a + b)},
$$
the shape parameters $a$ and $b$ depend on the conditional expectation and the conditional variance of $u_{k+1}$. When
$$
\mathrm{E}(u_{k+1} | u_k) = \phi_{34}(u_k),\quad \mathrm{Var}(u_{k+1} | u_k) = \dfrac{1}{1 + \eta_{34}} \phi_{34} (u_{34}) ( 1- \phi_{34}(u_{34})),
$$
then
$$
a = \eta_{34} \phi_{34}(u_k),\quad b = \eta_{34} (1 - \phi_{34}(u_k)).
$$
We can assume $\eta_{34}$ as a constant, and
$$\phi_{34}(u_k) = 0.08 + 1.06 \dfrac{(\alpha_{34} u_k)^{2.2}}{1 + (\alpha_{34} u_k)^{2.2}},$$
letting
$$\alpha_{34}  = 1.45 + 0.16\times [\mathrm{CD19}] + A_{34} \times \mathrm{Signal},$$
which represents the promotion of CD34 expression by CD19 and the CAR-T signal. For further details, please refer to \cite{Zhang:2021fs}.

Simulations presented in \cite{Zhang:2021fs} effectively replicated experimental results and predicted that CAR-T cell-induced cell plasticity could lead to tumor relapse in B-ALL after CD19 CAR-T treatment. 

\subsection{Cell-type transition mediated by epigenetic modifications}

Understanding how adult stem cells delicately balance self-renewal and differentiation remains a crucial equation in biological science. Here, we introduce a  hybrid model of stem cell regeneration based on the mathematical framework presented in this chapter. The model integrates a gene regulation network, epigenetic state inheritance, and cell regeneration, allowing for multi-scale dynamics from transcription regulation to cell population \cite{Huang:2024kg}.

The hybrid model contemplates a multi-cellular system, with each cell possessing a gene regulation network involving two genes that self-activate and repress each other. Cell regeneration behavior is modeled using a G0 cell cycle model, and the stochastic inheritance of epigenetic states during cell division is represented through the inheritance function. For a comprehensive understanding of the model, please refer to \cite{Huang:2024kg}.

Let $x_1$ and $x_2$ represent the expression levels of two genes, $X_1$ and $X_2$. The gene expression dynamics within one cell cycle can be modeled with ordinary differential equations:
\begin{equation}
\label{eq:g71}
\left\{
\begin{aligned}
\dfrac{\mathrm{d} x_1}{\mathrm{d} t} & = a_1 (\rho_1 + (1-\rho_1) \dfrac{x_1^n}{s_1^n + x_1^n}) + b_1 \dfrac{s_2^n}{s_2^n + x_2^n} - k_1 x_1,\\
\dfrac{\mathrm{d} x_2}{\mathrm{d} t} & = a_2 (\rho_2 + (1 - \rho_2) \dfrac{x_2^n}{s_2^n + x_2^n})  + b_2 \dfrac{s_1^n}{ s_1^n + x_1^n} - k_2 x_2,
\end{aligned}
\right.
\end{equation}
where $a_1$, $a_2$, $\rho_1$, $\rho_2$, $s_1$, $s_2$, $n$, $b_1$, $b_2$, $k_1$, and $k_2$ are non-negative parameters. Parameters $a_1$ and $a_2$ denote the maximum expression rates of self-activation of the two genes, while $b_1$ and $b_2$ are basal expression rates of the two genes without regression. Random fluctuations to the gene expression rates, for example, to $a_1$ and $a_2$, can be expressed as 
$$
a_i(\eta_i) = \alpha_i e^{\sigma_i \eta_i - \sigma_i^2/2}, \quad i=1,2,
$$ 
where $\alpha_1$ and $\alpha_2$ are positive parameters for the average expression rates. Here, $\sigma_1$ and $\sigma_2$ represent the intensities of noise perturbations, and $\eta_1$ and $\eta_2$ are colored noise defined by Ornstein-Uhlenbeck processes:
$$
\mathrm{d} \eta_i = -(\eta_i/\zeta_i) \mathrm{d} t + \sqrt{2/\zeta_i} \mathrm{d} W_i(t),\quad  i = 1, 2,
$$
where $W_1(t)$ and $W_2(t)$ are independent Wiener process, and $\zeta_1$ and $\zeta_2$ are relaxation coefficients. 

To incorporate the effects of epigenetic modification on gene regulation dynamics, it is known that epigenetic regulations, such as histone modification or DNA methylation, can interfere with chromatin structure that the expression levels $a_1$ and $a_2$ depend on the epigenetic modification states of the two genes, denoted by $u_1$ and $u_2$, respectively. The epigenetic states can refer to the fractions of marked nucleosomes or methylated CpG sites in a DNA segment of interest, and hence $\vec{u} = (u_1, u_2) \in \Omega = [0, 1]\times [0, 1]$. The epigenetic state primarily affects the chromatin structure and influences the chemical potential to initiate transcription. Thus, along with extrinsic noise perturbations, the expression rates $a_1$ and $a_2$ can be expressed as follows:
$$
a_i(u_i, \eta_i)  = \alpha_i e^{\lambda_i u_i} e^{\sigma_i \eta_i - \sigma_i^2/2}, i = 1, 2,
$$
where $\alpha_i (i=1,2)$ represents the impact of the epigenetic modification states on expression levels. 

The epigenetic states $u_1$ and $u_2$ undergo random changes only during cell division. Following the above argument, the random inheritance of the epigenetic state is represented by the inheritance function 
$$
p(\vec{u}, \vec{v}) = P(\mathrm{state\ of\ daughter\ cell}\ = \vec{u}\ |\ \mathrm{state\ of\ mother\ cell}\ = \vec{v}). 
$$
We assume that $u_1$ and $u_2$ vary independently during cell division, and hence,
$$
p(\vec{u}, \vec{v}) = p_1(u_1, \vec{v}) p_2(u_2, \vec{v}),
$$
where $p_i(u_i, \vec{v})$ represents the transition function of $u_i$, given the state $\vec{v}$ of the mother cell. Similar to the previous argument, we write the inheritance function $p_i(u_i, \vec{v})$ through the density function of beta-distribution, and assuming the conditional expectation and conditional variance of $u_i$ (given the state $\vec{v})$ as:
$$
\mathrm{E}(u_i | \vec{v}) = \phi_i(\vec{v}),\quad \mathrm{Var}(u_i | \vec{v}) = \dfrac{1}{1+\psi_i(\vec{v})} \phi_i(\vec{v}) (1 - \phi_i(\vec{v})).
$$
The function $\phi_i(\vec{v})$ and $\psi_i(\vec{v})$ together define the inheritance function. We assume that $\varphi_i(\vec{v})$ remains constant, while $\phi_i(\vec{v})$ increases with $v_i$ and is expressed by a Hill function as:
$$
\varphi_i(\vec{v}) = m_0, \phi_i(\vec{v}) = m_1  + m_2 \dfrac{(m_3 v_i)^{m_4}}{1 + (m_3 v_i)^{m_4}},\quad \vec{v} = (v_1, v_2), i = 1, 2,
$$
where $m_j$ $(j=0, 1, 2, 3, 4)$ are positive parameters.

The gene regulation network dynamics described above can be integrated with the G0 cell cycle model based on the dependence of kinetotypes on cell types. 

The gene regulation network dynamics given by \eqref{eq:g71} with properly selected parameter values can exhibit different patterns of steady states. Accordingly, we define the phenotype of cells following gene expressions of the steady states as follows: stem cells (SC) have medium expressions in both $X_1$ and $X_2$, transit-amplifying cells 1 (TA1) have high expression in $X_1$ and low expression in $X_2$,  transit-amplifying cells 2 (TA2) have low expression in $X_1$ and high expression in $X_2$, and transition cells (TC) otherwise. 

In the G0 cell cycle model, we only consider cells with the ability to undergo cell cycling, and each cell has different proliferation and cell death rates dependent on its cell phenotype. Cells that have lost the ability to experience cell cycling are removed from the system.

Different cell types differ in their cell proliferation regulation. For SCs, the proliferation rate can be given by 
$$
\beta_{\mathrm{SC}} = \beta_0 \dfrac{\theta}{\theta + Q(t)},
$$
where $Q(t)$ represents the number of SC at time $t$, $\beta_0$ represents the maximum proliferation rate, and $\theta$ is a constant for the half-effective cell number. TA cells, however, are assumed to have the maximum proliferation rate so that
$$
\beta_{\mathrm{TA1}} = \beta_{\mathrm{TA2}} = \beta_0.
$$
The removal rate $\kappa$, the apoptosis rate $\mu$, and the proliferation duration $\tau$ are assumed as constants. For details, please refer to \cite{Huang:2024kg}.

Given the epigenetic state $\vec{u} = (u_1, u_2)$ of each cell, the gene expression state $\vec{x} = (x_1, x_2)$ dynamically evolves according to the differential equations \eqref{eq:g71}. Accordingly, the cell phenotype and the kinetic rates $\beta$, $\kappa$, $\mu$, and $\tau$ can change during a cell cycle. When a cell undergoes mitosis, the cell divides into two cells, and the epigenetic states of the two daughter cells are calculated based on the inheritance functions. Simulation results in \cite{Huang:2024kg} demonstrate that random inheritance of epigenetic states during cell divisions can spontaneously induce cell differentiation, dedifferentiation, and transdifferentiation. Moreover, interfering with epigenetic modifications and introducing additional transcription factors can alter the probabilities of dedifferentiation and transdifferentiation, revealing the mechanism underlying cell reprogramming.

\section{Mathematical problems}
\label{sec:mp0}

This chapter introduces a mathematical framework \eqref{eq:6.26} of stem cell regeneration that considers cell heterogeneity and plasticity. This equation, a delay differential-integral equation, incorporates nonlocal transitions between different epigenetic states. Moreover, we further integrate stem cell regeneration dynamics with gene regulation network dynamics, leading to a novel population balance equation \eqref{eq:45}. This equation provides an integrative mathematical model describing the evolution of the population density $f(t, \vec{x})$, or Waddington's epigenetic landscape $U(t, \vec{x}) = - \log f(t, \vec{x})$.  While these equations provide novel mathematical formulations for quantifying the biological process of heterogeneous stem cell regeneration, many basic mathematical problems in understanding the biological processes remain open.

\subsection{Steady-state solution}
To consider the steady-state solution of \eqref{eq:6.26}, let $Q(t, \vec{x}) = Q(\vec{x})$ represent the steady state. The resulting equations are: 
\begin{equation}
\label{eq:m20}
\left\{
\begin{array}{rcl}
\displaystyle 0 &=&\displaystyle -Q(\vec{x}) (\beta(c, \vec{x}) + \kappa(\vec{x})) + 2 \int_\Omega \beta(c, \vec{y}) Q(\vec{y}) e^{-\mu(\vec{y})\tau(\vec{y})} p(\vec{x}, \vec{y}) \mathrm{d} \vec{y},\\
\displaystyle c &=&\displaystyle \int_\Omega Q(\vec{x}) \zeta(\vec{x}) \mathrm{d} \vec{x}.
\end{array}
\right.
\end{equation}
Substituting $c$ into the first equation, $Q(\vec{x})$ satisfies the nonlinear eigenvalue problem
\begin{equation}
\label{eq:E73}
\mathcal{L}_{c}[Q(\vec{x})] = 2 \int_\Omega \beta(c, \vec{y}) e^{-\mu(\vec{y})\tau(\vec{y})} p(\vec{x}, \vec{y}) Q(\vec{y}) \mathrm{d} \vec{y} - (\beta(c, \vec{x}) + \kappa(\vec{x})) Q(\vec{x}) = 0.
\end{equation}
Therefore, the problem of the existence and uniqueness of the steady-state solution reduces to finding a positive eigenvalue $c$ of the operator $\mathcal{L}_{c}$ such that the corresponding eigenfunction $Q(\vec{x})$ is non-negative for all $x\in \Omega$. When the eigenvalue $c>0$ exists, the solution of \eqref{eq:m20} can be obtained through rescaling the corresponding eigenfunction following the second equation in \eqref{eq:m20}.

Identifying the existence and stability of steady states is crucial for understanding the persistence of different biological states. While specific cases have been studied, open questions persist for general scenarios.

In the case of a finite discrete epigenetic state, and when the proliferation rate $\beta$ is independent of the epigenetic state, the steady-state problem has been discussed in \cite{Situ:2017dg}. Particularly, when $\beta$ is independent of the epigenetic state, the eigenvalue problem was reformulated as:
\begin{equation}
\mathcal{A}[Q] = \dfrac{1}{\beta(c)} Q,
\end{equation}
where $\mathcal{A}$ is a linear operator defined as
\begin{equation}
\mathcal{A}[Q] = \dfrac{1}{\kappa(\vec{x})}\left[2\int_{\Omega} e^{-\mu(\vec{y})\tau(\vec{y})}p(\vec{x}, \vec{y})Q(\vec{y}) \mathrm{d} \vec{y} - Q(\vec{x})\right].
\end{equation}
Thus, $(\beta(c))^{-1}$ is a positive eigenvalue of the operator $\mathcal{A}$. The existence of the eigenvalue can be obtained following the Perron-Frobenius theorem when there are a finite number of discrete epigenetic states \cite{Situ:2017dg}. 

The study in \cite{Situ:2017dg} further discussed the uniqueness and stability of the steady state when the inheritance function $p(\vec{x}, \vec{y})$ is independent of the state of the mother cell, i.e., $p(\vec{x}, \vec{y}) = p(\vec{x})$.

In the case of a 1-dimensional epigenetic state variable, when the delay $\tau > 0$ in the original equation \eqref{eq:6.26}, it may have oscillation solutions, and the positive steady state solution, if it exists, is unstable \cite{Liang:2023aa}. Conditions for oscillation solutions were studied in \cite{Liang:2023aa} through numerical simulations. Interestingly, the plasticity of cells during cell division can alter the requirements for oscillation solutions. However, the mathematical basis for the conditions for an oscillatory solution in the population number $Q(t, \vec{x})$ is unknown.

Moreover, the nonlinear eigenvalue problem \eqref{eq:E73} can be rewritten as:
\begin{equation}
\mathcal{L}_c[Q(\vec{x})] = \gamma(c) Q(\vec{x}),
\end{equation}
looking for the principle eigenvalue $\gamma$ for any parameter $c >0$. Here, the principle eigenvalue corresponds to $\gamma$ with a positive eigenfunction. Thus, if there is $c > 0$ such that $\gamma(c) = 0$, the problem \eqref{eq:E73} has a positive solution. The principle spectral theory has been applied \cite{Su:2023jg} to study the existence, uniqueness, and multiplicity of positive steady states to equations, as well as the long-time behavior of time-dependent solutions, when the proliferation rate $\beta$ is independent of the epigenetic state $\vec{x}$, and the delay $\tau$ is omitted in the original equation \eqref{eq:6.26}. Various explicit formulas for threshold values for tissue development, degeneration, and abnormal growth were obtained through the principle eigenvalue. For detailed discussions, please refer to \cite{Su:2023jg}.
  
\subsection{Entropy problem}

Based on the population density, the entropy of the multicellular system at time $t$ is defined as
\begin{equation}
\label{eq:Sp}
S(t) = -\int_{\Omega} f(t, \vec{x}) \log f(t, \vec{x}) \mathrm{d} \vec{x}.
\end{equation}
Thus, the derivative of the entropy $E(t)$ is given by 
\begin{equation}
\dfrac{\mathrm{d} S}{\mathrm{d} t} = - \int_{\Omega} \dfrac{\partial f(t, \vec{x})}{\partial t} \left(1 + \log f(t, \vec{x})\right) \mathrm{d} \vec{x}. 
\end{equation}
The entropy of a system measures the complexity of heterogeneity in the multicellular system, making the evolution of entropy biologically and mathematically interesting.

A study on abnormal growth induced by variations in microenvironmental conditions \cite{Huang:2024kg} revealed a nonlinear dependence of entropy changes on the cell population number $Q(t)$. As both $Q(t)$ and $S(t)$ are macroscopic measurements of the system, it is interesting to explore how we can formulate the macroscopic dynamics of the system during tissue growth.    

Following \cite{Qian:2001lk,Zhang:2012ke}, the derivative of the entropy for the stochastic differential equation \eqref{eq:sde1} is formulated as
\begin{equation}
\label{eq:ent1}
D \dfrac{\mathrm{d} S(t)}{\mathrm{d} t} = e_p(t) - h_d(t), 
\end{equation}
where $S(t) = -\int_{\Omega} P(t, \vec{x}) \log P(t, \vec{x}) \mathrm{d} \vec{x}$ is the entropy of the probability density function $P(t, \vec{x})$ at time $t$. Here, $e_p$ is the entropy production rate (EPR)
\begin{equation}
\label{eq:EPR1}
e_p(t) = \int_\Omega |\vec{F}(\vec{x}) - D \nabla \log P(t, \vec{x})|^2 P(t, \vec{x}) \mathrm{d} \vec{x},
\end{equation} 
and $h_d$ is the heat dissipation rate (HDR)
\begin{equation}
\label{eq:HDR1}
h_d(t) = \int_\Omega \vec{F}(\vec{x})\cdot \vec{J}(t, \vec{x}) \mathrm{d}\vec{x},
\end{equation}
with the probability flux $J(t, \vec{x})$ define by \eqref{eq:J0}. These formulas provide an interpretation of the gene circuit dynamics through statistical physics. 

Now, for the entropy \eqref{eq:Sp} associated with the population density $f(t, \vec{x})$, how can we formula the EPR and HDR? This question is crucial for understanding the statistical basis of tissue development as a non-equilibrium process, especially in the context of understanding entropy evolution during cancer development \cite{Hanselmann:2016kg,Tarabichi:2013ev}.

\subsection{Data-driven problem}

In the model framework \eqref{eq:6.26}, while the kinetic rates $\beta$, $\kappa$, $\mu$, and the proliferative duration $\tau$ can be obtained from cell regeneration dynamics, obtaining the inheritance function $p(\vec{x}, \vec{y})$ remains challenging with experimental data. Recent advancements in single-cell sequencing techniques have allowed quantification of gene expressions in individual cells. However, tracing cell division and quantifying molecular-level dynamics \textit{in vivo} remain challenging. Novel techniques for tracing individual cell division and single-cell lineages have been developed \cite{Quinn:2021aa,Denoth-Lippuner:2021aa}. We anticipate that these methods may eventually enable the measurement of the inheritance function $p(\vec{x}, \vec{y})$. 

Mathematically, addressing the following data-driven problem is crucial. Assuming we have obtained the population density $f(t, \vec{x})$ and the population size $Q(t)$ from experimental data, along with the knowledge of the kinetic rates $\beta$, $\kappa$, $\mu$, and the duration $\tau$, how can we derive the inheritance function $p(\vec{x}, \vec{y})$ based on the evolution equation \eqref{eq:fvh}? Furthermore, how can we derive the force term $\vec{F}$ in the population balance equation \eqref{eq:45}? Through these studies, we can gain insights into cell plasticity during cell regeneration, leading to an understanding of the dynamical Waddington landscape. A machine-learning method was proposed to construct a non-equilibrium potential landscape via a variational force projection formulation \cite{Zhao:2023arx}. This method offers an approach to dealing with the high-dimensional potential landscape. Additionally, a method for quantifying the pluripotency landscape of cell differentiation from single-cell RNA sequencing (scRNA-seq) data from continuous birth-death process was developed in \cite{Shi:2019gd}. This study provides a computational tool to quantify cell potency within the Waddington landscape based on scRNA-seq data.

\subsection{Local epigenetic state transition}

In the equations \eqref{eq:6.26} or \eqref{eq:6.26-1}, the inheritance function $p(\vec{x}, \vec{y})$ represents the non-local transition of epigenetic states during cell division. Here, we consider the situation when only local transition is allowed, \textit{i.e.}, 
\begin{equation}
\label{eq:m35}
p(\vec{x}, \vec{y}) = \varphi(\vec{y} - \vec{x})
\end{equation}
so that $\varphi(\vec{z}) > 0$ only when $\vert \vec{z} \vert < \epsilon$ for some $\epsilon > 0$. The function $\varphi(\vec{z})$ satisfies
\begin{equation}
\int_{\mathbb{R}^n} \varphi(\vec{z}) \mathrm{d} \vec{z} = 1,\quad \varphi(\vec{z}) \geq 0, \quad \forall \vec{z} \in \mathbb{R}^n.
\end{equation}
We assume the epigenetic state $\vec{x}\in \Omega \subset \mathbb{R}^n$. We further let
\begin{equation}
\alpha_i = \int_{\mathbb{R}^n} z_i \varphi(\vec{z}) \mathrm{d} \vec{z}, \quad D_{ij} = \int_{\mathbb{R}^n}  z_i z_j \varphi(\vec{z}) \mathrm{d} \vec{z}. 
\end{equation}

Now, we consider the equation \eqref{eq:6.26-1}, in which the delay $\tau$ is omitted. Substituting \eqref{eq:m35} into \eqref{eq:6.26-1}, we note that (here, we extend the integral over $\Omega$ to $\mathbb{R}^n$)
\begin{eqnarray*}
&&{}\int_{\mathbb{R}^n} \beta(c, \vec{y}) Q(t, \vec{y}) e^{-\mu(\vec{y})} p(\vec{x}, \vec{y}) \mathrm{d} \vec{y}\\
&=&  \int_{\mathbb{R}^n} \beta(c, \vec{x}) Q(t, \vec{x}) e^{-\mu(\vec{x})} \varphi(\vec{y} - \vec{x}) \mathrm{d} \vec{y}\\
&&{} + \int_{\mathbb{R}^n} (\vec{y} - \vec{x}) \cdot \nabla (\beta(c, \vec{x}) Q(t, \vec{x}) e^{-\mu(\vec{x})}) \varphi(\vec{y} - \vec{x}) \mathrm{d} \vec{y} \\
&&{} + \int_{\mathbb{R}^n} \dfrac{1}{2} (\vec{y} - \vec{x}) \cdot (\nabla^2 (\beta(c, \vec{x}) Q(t, \vec{x}) e^{-\mu(\vec{x})})) (\vec{y} - \vec{x}) \varphi(\vec{y} - \vec{x}) \mathrm{d} \vec{y}\\
&=& (1 + \sum_{i} \alpha_i \partial_i + \dfrac{1}{2} \sum_{i,j} D_{ij} \partial^2_{i,j})(\beta(c, \vec{x}) Q(t, \vec{x}) e^{-\mu(\vec{x})}),
\end{eqnarray*}
where
$$
\partial_i = \dfrac{\partial\ }{\partial x_i},\quad \partial_{i,j}^2 = \dfrac{\partial^2\ }{\partial x_i \partial x_j}.
$$
Thus, the equation \eqref{eq:6.26-1} is rewritten as
\begin{equation}
\label{eq:m38}
\left\{
\begin{aligned}
\dfrac{\partial Q(t, \vec{x})}{\partial t} &= (2\sum_{i} \alpha_i \partial_i  +  \sum_{i,j} D_{ij} \partial^2_{i,j})(\beta(c, \vec{x}) e^{-\mu(\vec{x})} Q(t, \vec{x}))\\
&\quad{} + ((2 e^{-\mu(\vec{x})} - 1) \beta(c, \vec{x})  - \kappa(\vec{x})) Q(t, \vec{x}) \\
c &= \int_\Omega Q(t, \vec{x}) \zeta(\vec{x}) \mathrm{d} \vec{x}.
\end{aligned}
\right.
\end{equation}
Particularly, if $D_{i,j}  = D \delta_{ij}$, we have a reaction-diffusion equation
\begin{equation}
\label{eq:m38'}
\left\{
\begin{aligned}
\dfrac{\partial Q(t, \vec{x})}{\partial t} &= \nabla\cdot(D \nabla + 2 \vec{\alpha})(\beta(c, \vec{x}) e^{-\mu(\vec{x})} Q(t, \vec{x}))\\
&\quad{} + ((2 e^{-\mu(\vec{x})} - 1) \beta(c, \vec{x})  - \kappa(\vec{x})) Q(t, \vec{x}) \\
c &= \int_\Omega Q(t, \vec{x}) \zeta(\vec{x}) \mathrm{d} \vec{x},
\end{aligned}
\right.
\end{equation}
where $\vec{\alpha} = (\alpha_1, \cdots, \alpha_n)$.

Equations \eqref{eq:m38} or \eqref{eq:m38'} describe the dynamics of stem cell regeneration with the local transition of epigenetic states during cell division. These are linear evolution equations, with global regulation involved through the growth factor $c$. The problems of steady-state solution, entropy, and data-driven analysis mentioned above can be formulated based on the local epigenetic state transition. 

\section{Discussions}

Cell division is a remarkable process in living organisms, where a parent cell divides into two daughter cells. These daughter cells can divide and grow independently, creating a new cell population from a single parental cell and its descendants. However, the two daughter cells may not be identical to their parental cell, leading to heterogeneity in the population of descendant cells due to the accumulation of epigenetic variations over multiple generations. The stable distribution of cell phenotypes forms the epigenetic landscape crucial in maintaining tissue homeostasis despite random cell loss and regeneration. Quantitative modeling of the mechanism underlying the dynamic equilibrium of the epigenetic landscape during tissue development is of great significance in understanding the rules governing living organisms \cite{Huang:2009dc,Brock:2009ke,Guillemin:2020aa,Davila-Velderrain:2015aa}.

The characteristics of a cell are determined by how genes are expressed and regulated within the cell. Differentiation equations can be used to model the biochemical reactions of the gene network, either deterministic or stochastic. However, differential equation models only apply during a single cell cycle and become invalid when the cell divides. During cell division, both the cell states and the model parameters may change abruptly, leading to diversity and adaptability in stem cell regeneration over time. Therefore, it is vital to incorporate gene network dynamics with cell division when quantitatively modeling long-term biological processes like tissue development and tumor progression.            

This chapter introduces two strategies for incorporating cell division into mathematical models of biological systems. The first strategy, Lagrange coordinate modeling, delves into the dynamics of gene networks within individual cells, resulting in random changes in cell states and model parameters during cell division. Each cell is represented by a unique set of equations, and the number of equations can change dynamically due to cell division. However, this approach makes the mathematical formulation challenging and lacks explicit inclusion of cellular behaviors such as proliferation, apoptosis, and differentiation in the model equations. To address these limitations, additional assumptions are necessary to describe the regulation of cell behaviors. Numerical studies of Lagrange coordinate models typically employ the agent-based modeling technique.

In contrast, the second strategy, Euler coordinate modeling, frames the evolution of population numbers of cells with the same epigenetic state through a differential-integral equation. This approach explicitly includes the regulations of cell behaviors, such as proliferation, apoptosis, and differentiation, in the equation. Plasticity resulting from cell division is represented as the inheritance probability function of epigenetic states, describing the conditional probability of epigenetic state changes in cell divisions. The inheritance function incorporates information on gene regulation networks. The Euler coordinate model integrates various biological interactions, encompassing the epigenetic state at the single-cell level, dynamic cell behavior, cytokine secretion, and the transition of epigenetic states. The concept of kinetotype is introduced through this model, offering a dynamic description of cell regeneration based on the epigenetic state of an individual cell \cite{Lei:2020fw}. Analogous to genotype, epigenotype, phenotype, kinetotype provides a crucial description of cell type, often associated with the activities of specific genes enriched in related pathways. However, defining the kinetotype from single-cell sequencing data remains challenging. 

These two modeling strategies are connected to different types of experimental data. Lagrange coordinate modeling describes molecular-level activities within individual cells, necessitating microscopic data for parameter identification. Such data include gene expression, epigenetic states, and protein levels associated with the genes of interest. On the other hand, Euler coordinate modeling focuses on the population-level dynamics of a multicellular system, requiring kinetic parameters related to cell renewal, differentiation, and cell death. Additionally, the heterogeneity and plasticity of cells are linked to single-cell sequencing data during the evolution process. Temporal single-cell RNA sequencing data can be crucial in determining how kinetic parameters may depend on the state of individual cells. However, the explicit dependencies are currently unknown and await further study.

The concept of Waddington's epigenetic landscape is crucial for deciphering the biological mechanisms that govern cell fate decisions and differentiation. This landscape is quantitatively characterized by the potential of epigenetic states, illustrated through the population density of cell states. Various modeling strategies exist to articulate the Waddington landscape, with one approach grounded in gene circuit dynamics featuring stochastic fluctuations in individual cells. This strategy leads to deriving the Fokker-Planck equation, aiming to model the evolution of population density. 

However, the Fokker-Planck falls short in addressing critical biological processes, such as the regulation of proliferation, the impact of cell division, and cell plasticity induced by these divisions. These processes play pivotal roles in non-equilibrium biological phenomena like embryo development and tumorigenesis, thereby making the Fokker-Planck equation potentially misleading in capturing their intricacies. 

Differing from the Fokker-Planck equation approach, the Euler coordinate model offers a more direct route to understanding the evolution of the Waddington landscape in the context of population dynamics during heterogeneous stem cell regeneration. This alternative approach derives an equation for the evolution of the population density of cells based on straightforward assumptions about cell proliferation, apoptosis, differentiation, and the transition of epigenetic states during cell division. In contrast to the Fokker-Planck approach, which delves into molecular-level dynamics inside a cell, the novel model equation \eqref{eq:fvh} accentuates cellular-level dynamics described by the kinetotype of cells. Kinetotypes, representing fundamental properties of the non-equilibrium processes in tissue development and tumorigenesis  \cite{Hanahan:2000hx,Lei:2020fw}, provide a more rational comprehension of complex biological processes. 

The Euler coordinate model has proven effective in studying diverse scenarios, including abnormal growth resulting from significant microenvironmental changes \cite{Zhang:2022vd}, the dynamic process of tumor cell plasticity-induced immune escape after CAR-T therapy in B-cell acute lymphoblastic leukemia (B-ALL) \cite{Zhang:2021fs}, and the dynamics of cell-type transitions mediated by epigenetic modifications \cite{Huang:2023hi}. In these scenarios, the evolution of cell population density induced by cell plasticity during division emerges as a critical determinant in achieving tissue homeostasis.

The Euler coordinate suggests a growth operator $\mathcal{R}$ for the population density, incorporating stem cell regeneration and plasticity during cell divisions. Applying the population balance analysis, we derive the dynamical equation \eqref{eq:45} for the population density $f(t, \vec{x})$. This equation comprises three key components: the term $-\nabla (f \vec{F})$ captures the gradient of cell states influenced by gene regulation networks, the diffusion term $\nabla\cdot(D \nabla f)$ represents fluctuation random molecular perturbations, and the growth operator $\mathcal{R}[f]$ accounts for stem cell regeneration and plasticity. While cell types are defined by states that vanish the gradient force $\vec{F}$, the diffusion term elucidates potential cell type switches independent of cell division. In contrast, cell plasticity coupled with cell division is embodied by the inheritance function $p(\vec{x}, \vec{y})$ within the growth operator. Thus, equation \eqref{eq:45} presents an integrated mathematical formulation harmonizing the dynamics of gene regulation networks with heterogeneous stem cell regeneration. 

In the Euler coordinate model equation \eqref{eq:6.26}, there is an ongoing debate on how to measure the epigenetic state of a cell. While single-cell RNA sequencing techniques provide high-dimensional data for a single cell, a low-dimensional variable is still crucial in characterizing the cell type and its regeneration kinetics. Biologically, several marked genes linked to proliferation, differentiation, and apoptosis signal pathways are reasonable candidates for quantifying the epigenetic states. However, a subset of marker genes cannot represent the overall epigenetic state of a cell, and the dimension of marker gene expressions remains large. 

Recently, machine learning methods have offered data-driven approaches to define macroscopic measurements of a cell based on single-cell RNA sequencing. These include the one-class logistic regression (OCLR) machine-learning method, which identifies the stemness feature \cite{Malta:2018ic}, the Mann-Whitney U statistic method, which evaluates the signature score of genes associated with a signaling pathway \cite{Andreatta:20212jg}, a computational framework (CytoTRACE) based on the number of expressed genes per cell \cite{Gulati:2020aa}, a computational framework (PhyloVelo) that estimates the velocity of transcriptomic dynamics by using monotonically expression genes through phylogenetic time \cite{Wang:2023aa}, and the single-cell entropy (scEntropy), which measures the order of cellular transcriptome profile \cite{Liu:2020de,Ye:2020fw}. These macroscopic measurements establish connections between the static single-cell RNA sequencing data and dynamic equations for the processes of tissue development. By integrating data-driven and model-driven studies, we can better understand complex biological processes, such as tissue development and tumor progression.

In summary, the Euler coordinate approach in the mathematical modeling of heterogeneous stem cell regeneration provides a reasonable understanding of the complex biological processes that govern the development of living organisms. The differential-integral equation model presented in this review offers a new perspective on the intricate problem of stem cell dynamics and a potential link between single-cell sequencing data and population dynamics.

\section*{Acknowledgments}
This work was supported by the National Natural Science Foundation of China (No. 12331018). 


\bibliographystyle{unsrt}


\end{document}